\documentstyle[psfig]{elsart}
\newcommand{\avg}[1]{\langle{#1}\rangle}
\newcommand{\Avg}[1]{\left\langle{#1}\right\rangle}
\newcommand{\be}{\begin{equation}}
\newcommand{\ee}{\end{equation}}
\newcommand{\beas}{\begin{eqnarray*}}
\newcommand{\eeas}{\end{eqnarray*}}
\newcommand{\bea}{\begin{eqnarray}}
\newcommand{\eea}{\end{eqnarray}}
\newcommand{\req}[1]{(\ref{#1})}
\newcommand{\ovl}[1]{\overline{#1}}
\def\sign{\hbox{sign}\,}
\def\Tr{\hbox{Tr}}
\def\nit{\hbox{\it I\hskip -2pt  N}}
\def\A{{\cal A}}
\def\P{{\cal P}}
\def\N{{\cal N}}
\def\AP{{\cal A}^{\cal P}}
\def\DN{\Delta^{\cal N}}
\def\va{{\vec a}}
\def\vp{{\vec \pi}}
\def\vz{{\vec z}}
\def\va{{\vec a}}
\def\rit{\hbox{\it I\hskip -2pt  R}}

\begin{document}

\begin{frontmatter}
\title{Exact solution of a modified El Farol's bar problem:
Efficiency and the role of market impact}
\thanks{We acknowledge J. Berg, S. Franz and Y.-C. Zhang for 
discussions and useful suggestions. On the economic side we 
profited greatly from critical discussions with A. Rustichini and 
F. Vega-Redondo on learning and dynamic games.
This work was partially supported by Swiss National Science 
Foundation Grant Nr 20-46918.98.}
\author{Matteo Marsili}
\address{Istituto Nazionale per la Fisica della Materia (INFM),
Trieste-SISSA Unit,\\
V. Beirut 2-4, Trieste I-34014}
\author{Damien Challet}
\address{Institut de Physique Th\'eorique, 
Universit\'e de Fribourg, CH-1700}
\author{and Riccardo Zecchina}
\address{The Abdus Salam International Centre for Theoretical
Physics Strada Costiera 11, P.O. Box 586, I-34014 Trieste}
\date{\today}

\begin{abstract}
We discuss a model of heterogeneous, inductive rational agents 
inspired by the El Farol Bar problem and the Minority Game. 
As in markets, agents interact through a collective aggregate
variable -- which plays a role similar to price -- whose value is
fixed by all of them. Agents follow a simple reinforcement-learning 
dynamics where the reinforcement, for each of their available
strategies, is related to the payoff delivered by that strategy.
We derive the exact solution of the model in the 
``thermodynamic'' limit of infinitely many agents using
tools of statistical physics of disordered systems. Our results 
show that the impact of agents on the market price plays 
a key role: even though price has a weak dependence on the
behavior of each individual agent, the 
collective behavior crucially depends on whether agents account 
for such dependence or not. Remarkably, if the adaptive behavior
of agents accounts even ``infinitesimally'' for this dependence they
can, in a whole range of parameters, reduce global fluctuations by
a finite amount. Both global efficiency and individual utility
improve with respect to a ``price taker'' behavior if agents 
account for their market impact. 
\end{abstract}
\end{frontmatter}

\newpage
%\tableofcontents

\section{Introduction}

The El Farol bar problem \cite{Arthur} has become a popular
paradigm of complex systems. It describes the situation where 
$N$ persons have to choose whether to go or not to a bar which is
enjoyable only if it is not too crowded. In order to choose,
each person forms mental schemes, hypotheses or behavioral rules
based on her beliefs  and she adopts the most successful one on
the basis of past performance. Inductive \cite{Arthur}, 
low \cite{roth} or generally bounded rationality based on learning
theory \cite{learn} is regarded as a more realistic approach to 
the behavior of real agents in complex strategic situations\cite{rust2}. 
This is specially true in contexts involving many heterogeneous 
agents with limited information, such as the  El Farol bar problem.
Theoretical advances, beyond numerical simulations, is 
technically very hard on these problems and it has been regarded 
as a major step forward in the understanding of complex 
systems\cite{Casti}. 

The minority game \cite{CZ1,ZENews} represents a first step in
this direction. It indeed describes a system of interacting agents
with inductive rationality which face the problem of finding
which of two alternatives shall be chosen by the minority.
This problem is quite similar in nature to the El Farol bar 
problem as the result for each agent depends on what all other
agents will do and there is no {\em a priori} best alternative. These
same kind of situations arise generally in systems of many 
interacting adaptive agents, such as markets\cite{ZENews,Savit}.

Numerical simulations by several authors 
\cite{CZ1,Savit,johnson,cavagna,CM,comexpMG} have 
shown that the minority game (MG) displays 
a remarkably rich 
emergent collective behavior, which has been qualitatively 
understood to some extent by approximate schemes
\cite{ZENews,CZ2,johnscrowd}. 
In this paper, which follows refs. \cite{CM,CMZ}, we study a generalized
minority game and show that a full statistical characterization of its
stationary state can be derived analytically in the ``thermodynamic''
limit of infinitely many agents. Our approach is based on tools 
and ideas of statistical physics of disordered systems\cite{MPV}.

The minority game, as the El Farol bar problem, allows for a 
relatively easy definition in words. This may be enough for
setting up a computer code to run numerical simulations, but
it is clearly insufficient for an analytical approach.
Therefore we shall, in the next section, define carefully
its mathematical formulation. We shall only
discuss briefly its motivation, for which we refer the 
reader to refs. \cite{CZ1,ZENews,Savit}. Even though the behavioral 
assumptions on which the MG is based may be questionable when applied to 
financial markets (see sect. \ref{motivation}), still we find it 
convenient to consider and discuss the model as a toy model for a market, 
in line with refs. \cite{ZENews,Savit,prodspec}. The relation to markets, 
at this level, may just be seen as a convenient language to discuss the 
results in simple terms. This choice reflects our taste and surely more work 
needs to be done to show the relation of the minority game with real 
financial markets. We believe, however, that because of the statistical 
nature of the collective behavior -- which are usually quite robust with
respect to microscopic changes -- our results may be qualitatively 
representative of generic systems of agents interacting through a global 
quantity {\em via} a minority mechanism, such as markets. 
The minority game indeed captures the essential interaction between
agents beliefs and market fluctuations -- how individual beliefs, processing 
fluctuations, produce fluctuations in their turn. This interaction is usually
shortcut in mathematical economy assuming market efficiency, i.e. that 
prices {\em instantaneously} react to and incorporate agents beliefs. 
The motivation underlying the efficient market hypothesis, is, in few
words that if there where inefficiencies -- or arbitrage opportunities --
that would be exploited by speculators in the market and washed out
very quickly. Implicitly one is assuming that there is an infinite
number of agents in the market who are using very sophisticated
strategies which can detect, exploit and eliminate arbitrages very quickly.
As ``stylized'' as it may be, the minority game allows to study how a finite 
number of heterogeneous agents interact in a complex system such as a market.
It allows to ask to what extent this ``stylized'' market is inefficient and
how agents really exploit arbitrage opportunities and to what extent.

After defining the stage game, we shall briefly discuss its Nash 
equilibria: these are the reference equilibria of deductive rational 
agents. Finally we shall pass to the repeated game with
adaptive agents which follow exponential learning. 
We show that the key difference between agents 
playing a Nash equilibrium and agents in the usual minority game
is {\em not} that the first are deductive whereas the latter are 
inductive. Rather the key issue is whether agents account for 
their ``market impact'' or not. By market impact we refer to the fact that
the choice of each agent affects aggregate quantities, such as prices.
In the minority game \cite{ZENews,Savit} agents behave as ``price takers'', 
i.e. as if their choices did not affect the aggregate. 
However, due to the minority nature of the interaction,
the market impact reduces the ``perceived'' performance of strategies 
which agents use in the market with respect to those which 
which are not used and whose performance is monitored on the basis
of a virtual trade (assuming the same price).
In order to analyze in detail this issue, we generalize the MG and 
allow agents to assign an extra reward $\eta$ to a strategy when it is 
played. This parameter allows agents to account for their market 
impact and it plays the same role as the Onsager reaction term, or
cavity field, in spin glasses\cite{MPV}. 

Our main results are:
\begin{enumerate}
\item We derive a continuum time limit for the dynamics of learning.
\item We show that this dynamics admits for a Lyapunov function, i.e.
a function of all relevant dynamical variables which decreases on all
trajectories of the dynamics. This is a very important result since
it turns the problem of studying the stationary state of a stochastic
dynamical system into that of characterizing the (local) minima of 
a function. Considering this function as an Hamiltonian, we can
apply the tools of statistical mechanics to solve the problem.
\item When agents do not consider their impact on the market, as in the
minority game ($\eta=0$), {\em i)} the stationary state is unique.
{\em ii)} the Lyapunov function is a measure of the asymmetry of the market. 
In loose words, agents minimize market's predictability. 
{\em iii)} When the number of agents exceeds a critical number the market
becomes symmetric and unpredictable, with large fluctuations as
first observed in \cite{Savit}.
\item If agents know what is the dependence of the aggregate variable
on their behavior they can consider their impact on the market. We refer
to this as the {\em full information} case since agents have full information 
on how the aggregate would have changed for each of their choices.
In this case, {\em i)} there are exponentially many stationary states. 
{\em ii)} these states are Nash equilibria, {\em iii)} the Lyapunov function 
measures market's fluctuations, which means that agents cooperate optimally 
in maximizing global wealth when maximizing their own utility. As a result, 
fluctuations always decreases as the number of agents increase.
\item This state is recovered when $\eta=1$. This means that 
agents need not have full information in order to reach this optimal
state. It is enough that they over-reward the strategy they are currently
playing with respect to those they are not playing, by a quantity $\eta$.
\item Any $\eta>0$ implies an improvement both in individual payoffs --
as shown in sect. \ref{agentview} and in 
global efficiency with respect to the $\eta=0$ case.
\item The most striking result comes when asking how does the collective 
behavior interpolates between the two quite different limits when
changing $\eta$ from $0$ to $1$. The result is that when there are few 
agents the change is mild and continuous -- even though there is a phase 
transition, that is a continuous one (second order). When there are many 
agents the change happens suddenly and discontinuously as soon as $\eta>0$. 
Even an infinitesimal $\eta$ is enough to reduce market's fluctuations
by a finite amount. 
\end{enumerate}

These results suggests that the neglect of market impact -- which seems an 
innocent approximation\footnote{The impact of each agent on the aggregate 
is of relative order $1/N$ and it vanishes as $N\to\infty$.} and is usually 
at the very basis of
mathematical economy and finance\footnote{For example, in determining
optimal investment strategies or pricing, it is customary to consider
prices as just exogenous processes, independent of the trading strategy
really adopted.} -- plays a very important role in complex systems such
as markets.

\section{The stage game: strategic structure}

\subsection{Actions and payoffs}

The minority game describes a situation where a large number
$N$ of agents have to make one of two opposite actions -- such as
e.g. ``buy'' or  ``sell'' -- and only those agents who choose the
minority action are rewarded. This is similar to the El
Farol bar problem, where  each one of $N$ agents may either
choose to go or not to a bar which  is enjoyable only when it is
not too crowded. In order to model this situation, 
let $\N=(1,\ldots,N)$ be the set of agents and let $\A=(-1,+1)$
be the set of the two possible actions. If $a_i\in\A$ is
the action of agent $i\in\N$, the payoffs to agent $i$ is
given by
\be
u_i(a_i,a_{-i})=-a_i\,A~~~\hbox{ where }~~~A=\sum_{i\in\N} a_i,
\label{ui0} 
\ee
where $a_{-i}=\{a_j,\,j\ne i\}$ stands for opponents actions. The
game rewards the minority group.
To see this, note that the total payoff to 
agents $\sum_i u_i=-A^2$ is always negative. Then the majority of 
agents, who have $a_{i}=\sign A$, receives a negative payoff
$-|A|$, whereas the minority  ``wins'' a payoff of $|A|$. Eq. \req{ui0}
can be generalized to $u_i(a_i,a_{-i})=-a_i\,U(A)$: if the 
function $U(x)$ is such that $x\,U(x)=-xU(-x)\ge 0$  for all 
$x\in \rit$, the game again rewards the minority. The
original model\cite{CZ1,ZENews} takes $U(x)=\sign x$, but the
collective behavior is qualitatively the same \cite{CM} as that of 
the linear case $U(x)=x$ on which we focus. Note that the
``inversion'' symmetry $u_i(-a_i,-a_{-i})=u_i(a_i,a_{-i})$ implies 
that the two actions are {\em a priori} equivalent:
there cannot be any best actions, because otherwise everybody
would do that and loose.

The key issue, clearly, is that of coordination. 
With respect to coordination games \cite[chapt. 6]{coordgames},
we remark that agents cannot communicate. If communication were 
possible, agents would have
incentives to stipulate contracts -- such as ``We
toss a coin, if the outcome is head I do $a_{\rm me}=+1$ and you do
$a_{\rm you}=-1$, and if it is tail we do the other way round''. 
Both players would benefit from this contract because it transforms
the negative sum game into a zero sum game for the two players. The
contract would then be self-enforcing. 

Agents interact only through a {\em global} or {\em aggregate}
quantity $A$ which is produced by all of them. This type of 
interaction is typical of market systems \cite{ZENews} ant it
is similar to the long-range interaction assumed in mean-field
models of statistical physics\cite{MPV}. Finally note that the
El Farol bar problem has a similar structure but with $A$
replaced by $(A-A_0)$ in Eq. \req{ui0} where $A_0$ is related to 
the bar's comfort level \cite{Arthur,johnsbar}.

\subsection{States and information}

Nature can be in one of $P$ {\em states}, which are labelled by a
variable $\mu$ which takes integer values 
$\mu\in{\cal P}\equiv\{1,2,\ldots,P\}$. We assume that 
$P$ is large and of the same order of $N$ and we define
$\alpha\equiv P/N$, which we eventually keep finite in the 
limit $N\to\infty$. The reason for this particular limit 
is because, as first shown in ref. \cite{Savit}, the model's 
behavior only depend on the combination $\alpha=P/N$ in the large 
$N$ limit. 
The variable $\mu$ encodes all possible information 
on the state of the environment where agents live, so we shall
sometimes call $\mu$ ``information''. 
$\mu$ is drawn from a distribution $\varrho^\mu$ on 
${\cal P}$, independently at any time step. Most results will
be presented for the uniform case $\varrho^\mu=1/P$.
In both the El Farol 
model and in the minority game, $\mu$ has a different, more complex
definition, on which we shall return later in section \ref{endoexo}.
In what follows, we shall denote with an over-line 
$\ovl{O}\equiv \sum_{\mu\in\P} \varrho^\mu O^\mu$ the 
average of a quantity $O^\mu$ over $\mu$. 
For any $\mu\in{\cal P}$, payoffs are still given by
Eq. \req{ui0}. Strictly speaking $\mu$ is a so called
{\em sun-spot}\cite{sunspot} because the payoffs only depend on the
actions of agents. Now, however, the {\em pure} strategies of each
agent may depend on $\mu$. We call $\AP$ the set of all such 
strategies: An element of $\AP$ is a function $a:\mu\in\P\to a^\mu\in 
\A$ or a $P$ dimensional vector with coordinates $a^\mu$,
$\forall\mu\in \P$\footnote{We use the simple letter $a$ without
the index ${}^\mu$ to denote the function.}.
There are $|\AP|=2^P$ possible such functions. 
We call $a_i\in\AP$ a possible pure strategy for agent $i\in\N$,
with elements $a_i^\mu\in\A$ for all $\mu\in\P$.
With this notations, the payoff ``matrix'' reads 
$\ovl{u_i}(a_i,a_{-i})=-\ovl{a\,A}$ where $A^\mu=\sum_i a_i^\mu$.
At this level we have just replicated the game of the previous 
level $P$ times. Again there cannot be a best strategy $a\in\AP$ 
for the same reasons as before. 

\subsection{Heterogeneous beliefs and strategies}
\label{gamedef}

Now we assume that each agent only restricts her choice on a small
subset of $S$ elements of $\AP$. 
We use the vector
notation $\va_i=(a_{1,i},\ldots,a_{S,i})$ to denote the subset
of strategies available to agent $i$, with elements 
$a_{s,i}\in\AP$. The action of agent $i$, when the state
is $\mu$ and she chooses her $s^{\rm th}$ strategy shall 
then be $a_{s,i}^\mu\in\A$. We shall mainly work on the case where
the strategies $a_{s,i}$ are randomly and
independently drawn (with replacement) from $\AP$. More precisely
\be
P(a_{s,i}^\mu=+1)=P(a_{s,i}^\mu=-1)=\frac{1}{2},~~~\forall 
i\in \N,~s\in \{1,\ldots,S\},~\mu\in\P.
\label{probasim}
\ee
Note that independence of $\va_i$ across agents is reasonable because
$\mu$ is a sunspot and no pre-play communication is possible (agents
are assigned their $\va_i$ before the game starts). 

The utility of agent $i$, given 
the value of $\mu$, his choice $s_i$ and the choice of other agents
$s_{-i}=\{s_j,~j\neq i\}$, now becomes
\be
u_i^\mu(s_i,s_{-i})=-a^\mu_{s_i,i}\, A^\mu
\hbox{~~~~with~~~~}
A^\mu\equiv\sum_{j\in\N} a_{s_j,j}^{\mu}.
\label{ui}
\ee
The goal of each agent is to maximize his expected payoff
over all possible values of $\mu$, which, for agent $i$, reads
\be
\ovl{u_i}(s_i,s_{-i})=-\ovl{a_{s_i,i}\,A}\equiv
-\sum_{\mu=1}^P \varrho^\mu a_{s_i,i}^\mu\,A^\mu
\label{ui2}
\ee

\subsection{Motivation}
\label{motivation}

This structure was introduced in refs. \cite{CZ1,ZENews} in order
to model {\em inductive rational} behavior of agents \cite{Arthur}.
In few words the argument is the following:
If agents have not a completely detailed model of the game they are engaged
in, they may think that the value $\mu$ has some  effect on the
game's outcome $A$, eventually because she believes that other
agents will believe the same. This is a self-reinforcing belief
because if agents behave differently for different values of $\mu$,
the aggregate outcome $A$ will indeed depend on $\mu$, thus
confirming agents' beliefs. The game's structure, however, is such
that there can be no ``commonality of expectations'' \cite{Arthur}.
This means that,  since there is no rational best strategy $a_{\rm best}$
-- because otherwise everybody would use that and loose --
agents' expectations (or beliefs) are forced to differ. 

On the basis of her mental schemes or hypotheses on the situation she 
faces, an agent may consider a particular strategy more ``likely'' to 
predict the ``correct'' action than another one. More precisely, she 
may consider that only the $S$ forecasting rules $\va_i=(a_{1,i}\ldots,
a_{S,i})$, out of all the $2^P$ such rules in $\AP$, are ``reasonable''
or compatible with her beliefs and then restrict her choice to 
just those. Heterogeneous beliefs are represented by the fact that
each agent draws her strategies at random, independently from others
using Eq. \req{probasim}. Alternatively, one may think, following 
Aumann \cite{Aumann},
that the $\va_i$ are ``rules of thumb'' that agent $i$ evolved or learned
in other contexts and which she applies in this context.
We refer the reader to refs. \cite{Arthur,Aumann} 
for a deeper discussion of this behavior. 

It is worth to remark that, in this view, the restriction from $2^P$ to
$S$ strategy is voluntary. It is not difficult to argue that agents have
payoff incentives to increase the number of their strategies. 
In refs. \cite{Arthur,CZ1,ZENews}, the state $\mu$ is known to agents 
before they take a decision. So why don't agents take a decision
$s_i^\mu$ which depends on $\mu$ -- or even an action $a_i^\mu$ which
depends on $\mu$? The answer is that they do not do so because of computational
costs one is implicitly building in the model: We are assuming that agents, 
in complex strategic situations, prefer to simplify decision tasks.

While this motivation may seem reasonable for an issue such as going to
a bar or not \cite{Arthur}, 
it may not be appropriate for agents in financial markets
as proposed in ref. \cite{ZENews}. 

On the other hand, this game's structure is justified if we assume that
agents do not know the state $\mu$ before their choice. Indeed if 
$\va_{i}^\mu$ where known to agent $i$ before taking her
decision, it would be reasonable for her to decide her best action 
conditional on her {\em private} information $\va_{i}^\mu$. 
If $\mu$ is not known in advance, one can imagine a situation where 
agents resort to $S$ 
``devices'' which take actions $a_{s,i}^\mu$ for them. The restriction 
in strategy space in this case is due to some implicit constraints
or costs.

Rather then defending the behavioral assumptions of refs. 
\cite{Arthur,CZ1,ZENews}
or pushing further these arguments, we prefer to remain at a generic level.
Indeed we believe the model displays collective phenomena which are of a 
generic relevance, because of their statistical nature. 
Furthermore, this collective behavior is so rich that, in our opinion, 
it deserves investigation by its own.

\subsection{Notation: mixed strategies and averages}

Before coming to the analysis of the game, let us introduce 
{\em mixed strategies}. The mixed strategy $\pi_{s,i}$, $s=1,\ldots,S$ 
of each agent $i$ is a distribution over her available strategies:
$\pi_{s,i}$ is, in other words, the probability with which 
agent $i$ plays her $s^{\rm th}$ strategy. Again we use a vector notation
$\vp_i=(\pi_{1,i},\ldots,\pi_{S,i})\in\Delta_i$, where 
$\Delta_i$ is the $S$ dimensional simplex of $i^{\rm th}$
agent. We introduce 
the scalar product $\vec{u}\cdot\vec{v}\equiv \sum_{s=1}^S
u_s, v_s$ for vectors $\vec{u},\vec{v}\in\rit^S$. We also
define the norm $|v|^2\equiv\vec{v}\cdot\vec{v}$ of vectors in
$\rit^S$. Expectations on the mixed strategy of player $i$ reads 
$E_{\vp_i}(\vec v)=\vp_i\cdot\vec{v}$. We also define the 
direct product $\Delta^\N=\prod_{i\in \N}\Delta_i$ which we shall
also call the {\em phase space}. A point $(\vp_1,\ldots,\vp_N)
\in\Delta^\N$ is indeed a possible state of the system. 
Finally we use the shorthand notation 
\be
\avg{O}=\sum_{s_1=1}^S\ldots\sum_{s_N=1}^S\pi_{s_1,1}\cdots
\pi_{s_N,N}\ O_{s_1,\ldots,s_N}
\ee
for the expectation on the product measure of mixed strategies 
over the phase space $\DN$. For example we shall frequently
refer to the quantity
\be
\Avg{A^\mu}=\sum_{i\in\N}\vp_i\cdot\va_i^\mu.
\label{avgAmu}
\ee

\section{Characterization of collective behavior}

As a preliminary to a more detailed discussion, we find it useful
to introduce the key quantities which describe the collective
behavior. First, as a measure of global efficiency, we take
\be
\sigma^2\equiv
\ovl{\Avg{A^2}}=\sum_{\mu\in\P}\varrho^\mu\Avg{\left(\sum_{i\in\N}
a_{s_i,i}^\mu\right)^2}
\ee
where we remind that $\avg{\ldots}$ means expectation 
over the variables $s_i$ with the corresponding mixed strategy
distribution $\pi_{s_i,i}$. Note that 
$\sigma^2=-\sum_i\avg{\ovl{u_i}}$ is just the total
loss of agents. A small value of $\sigma^2$ implies an
efficient coordination among agents.

By construction, the model is symmetric in the sense that, for any
$\mu$, no particular sign of $A^\mu$ is {\em a priori} preferred. 
We shall see, however, that this symmetry can be ``broken''
resulting in a state where $A^\mu$ may take more probably positive
than negative values for some $\mu$ and {\em vice-versa} for other
values of $\mu$. As a measure of this asymmetry, it is useful to
introduce the quantity
\be
H\equiv\ovl{\Avg{A}^2}=\sum_{\mu=1}^P\varrho^\mu\Avg{A^\mu}^2
=\sum_{\mu=1}^P\varrho^\mu\left(\sum_{i\in\N}\vp_i\va_i^\mu\right)^2.
\ee
Note that $H>0$ implies that there is a {\em best} strategy 
$a^\mu_{\rm best}=-\sign\avg{A^\mu}$ that could ensures a
positive payoff to a new-comer agent. Ideally
because if the new-comer really starts playing the game, she will
also affect the outcome $A^\mu$. This suggests that $H$ can be
regarded as a measure of the exploitable {\em information
content} of the system by an external agent\cite{CM}.

Both of these quantities are {\em extensive}, i.e. are
proportional to $N$ for $N\to\infty$, and we shall mainly be
interested in the finite quantities $\sigma^2/N$ and $H/N$.
As a statistical characterization of the equilibrium, it is useful 
to introduce the {\em self overlap}
\be
G=\frac{1}{N}\sum_{i\in \N}|\pi_{i}|^2
=\frac{1}{N}\sum_{i\in \N}\sum_{s=1}^S\pi_{s,i}^2
\label{G}
\ee
which gives a measure of the average spread of mixed strategies played
by agents. If all agents play pure strategies $G=1$ whereas $G=1/S$ if
$\pi_{s,i}=1/S$ $\forall s,i$. Therefore $1/G$ is a measure of the
``effective'' number of strategies that agents play on average.
Note that $\sigma^2$ can be written as
\be
\frac{\sigma^2}{N}=\frac{H}{N}+1-G-\frac{1}{N}
\sum_{i\in\N}\sum_{s, s'\neq s}\pi_{s,i}\pi_{s',i}\ovl{a_{s,i}a_{s',i}}
\cong\frac{H}{N}+1-G
\label{sigmaH}
\ee
where in the last relation we neglected terms which vanish in the limit 
$N\to\infty$ (because $\ovl{a_{s,i}a_{s',i}}\sim P^{-1/2}$ for $s\neq s'$).
Eq. \req{sigmaH} means that the loss of agents come either the asymmetry 
$H$ which they produce or from the stochastic fluctuations of their choices.
Indeed if agents play pure strategies, $G=1$ and the last term vanishes.
Put differently, the stochastic fluctuations $\sigma^2$ of the market --
or volatility -- has a systematic contribution $H$ arising from unexploited
asymmetries and a stochastic one $1-G$, which is generated by stochastic
choice of agents.

%It is also of interest to consider the distribution of mixed strategies
%across the population. In particular one may ask what is the 
%fraction $\phi_k$ of agents that use only $k$ out of the $S$ strategies 
%at her disposal
%\be
%\phi_k=\frac{1}{N}\sum_{i=1}^N\delta_{k,k_i},~~~~~k_i=\sum_{s:\pi_{s,i}>0} 
%1,
%\ee
%where $\delta_{i,j}=0$ if $i\ne j\in\nit$ and $\delta_{i,i}=1$ for
%$i\in\nit$.

\section{Nash Equilibria}
\label{Nash}

Given the payoffs and the choices available to agents we now briefly discuss 
Nash equilibria of the stage game. The motivation for this section is that, 
on one hand Nash equilibria shall provide a reference framework for the 
following discussion. On the other this discussion allows to appreciate the
strategic complexity of the problem. For our purposes, Nash equilibria 
are those states which are stable under payoff incentives, given the choices
available to agents. We shall not discuss refinements. We shall remain at
a quite simple level without pretending either completeness or rigor.

Given the symmetry of the game, let us first look
for symmetric Nash equilibria at the level of actions only (i.e. $P=1$) with
no restriction on strategies ($\va_i=\A$ for all $i\in\N$). 
These cannot be in pure actions so let us look for Nash equilibria in mixed
actions: Each player either plays
$a_i=+1$ with probability $\pi_i$ or she plays $a_i=-1$ otherwise.
It is easy to see that $\pi_i=1/2$ is the only symmetric Nash
equilibrium: No agent has incentives to deviate from the choice 
$\pi_i=1/2$ if others stick to it. This state, which we shall call the
{\em random agent state}, is usually taken as a reference
state\cite{Savit,CZ2} and it is characterized by
$\sigma^2=N$ and $H=0$. 

The game has many more Nash equilibria than the symmetric one. For example
in pure actions, any state where $|A|=1$ and $N$ odd (or $|A|=0$ and
$N$ even) is a Nash equilibrium. Indeed, focusing on $N$ odd, agents in 
the minority (playing $a_i=-A$) would decrease her payoff, switching to the 
majority side. On the other hand agents in the majority cannot increase 
their payoff
changing from $a_i=A$ to $a_i=-A$ because then, also the majority
would change $A\to -A$. The number $\Omega_{\rm Nash}^{(0)}$ of 
these Nash equilibria is 
\be
\Omega_{\rm Nash}^{(0)}={N\choose \frac{N-1}{2}}+{N\choose
\frac{N+1}{2}},\qquad\qquad {N\choose k}\equiv\frac{N!}{k!(n-k)!}
\ee
which is exponentially large in $N$.
Each of these states is globally optimal, since it has no
fluctuation: $\sigma^2=H=1$ as compared to $\sigma^2=N$, $H=0$ in the 
symmetric Nash equilibrium.

There are many more Nash equilibria at this simple 
level\footnote{Consider e.g. 
to split the population $N$ into two groups of $K$ agents playing pure
actions and $N-K$ playing symmetric mixed actions.}. 
These simple considerations are already enough to appreciate the complexity
of the problem. Note in particular that virtually all possible collective
behavior, as parametrized by $\sigma^2$ and $H$ is possible.

The complexity increases when $P>1$ and agents have still 
no restriction on strategies ($\va_i=\AP$ for all $i\in\N$)
Again we have the symmetric Nash equilibrium --
the random agents state -- where, for any $\mu$, each agent plays
$a=+1$ with probability $1/2$. The number of Nash
equilibria in pure actions is huge. Indeed any combination
of $P$ pure actions Nash equilibria at the level of actions 
(one for each value of $\mu$), is a Nash equilibrium. There are therefore
$\Omega_{\rm Nash}^{(1)}=
\left(\Omega_{\rm Nash}^{(0)}\right)^P$ such equilibria each of
them with minimal fluctuations $\sigma^2=1$. Again there are many more
Nash equilibria. 

When the strategies of agents are restricted to the sets $\va_i$ and
$S$ is small (typically finite when $N,P\to\infty$), the problem of
identifying Nash equilibria becomes more complex. One way to tackle
the problem is to write down the multi-population standard replicator 
dynamics \cite{evol} and then
identify Nash equilibria in evolutionarily stable strategies by its 
stationary and stable points. Again we make no claim of completeness:
We just focus on a particular subclass of Nash equilibria\footnote{These
are {\em strict} Nash equilibria.} -- 
which are evolutionarily stable -- which shall
play a peculiar role in the following.

The multi-population standard replicator 
dynamics \cite{evol} (RD) reads
\be
\frac{d\pi_{s,i}}{dt}=-\pi_{s,i}\sum_{j\in\N,\,j\ne i}
\left[\ovl{a_{s,i}(\vp_j\cdot\va_j)}
-\ovl{(\vp_i\cdot\va_i)(\vp_j\cdot\va_j)}\right].
\label{RD}
\ee
We observe that $\sigma^2=\sum_{i, j\ne i}\ovl{
(\vp_i\cdot\va_i)(\vp_j\cdot\va_j)}+N$ is a Lyapunov function 
under this dynamics. Indeed a little algebra leads to
\be
\frac{d\sigma^2}{dt}=-2\sum_{i\in \N}\sum_{s=1}^S
\pi_{s,i}
\left[
\ovl{(a_{s,i}-\vp_i\cdot\va_i)\sum_{j\ne i}\vp_j\cdot\va_j}
\right]^2\le 0.
\label{lyapsig}
\ee
Therefore Nash equilibria are local minima of $\sigma^2$ in $\DN$.
Furthermore, $\sigma^2$ is a linear function of $\pi_{s,i}$ for 
any $i,s$, so that $\frac{\partial^2\sigma^2}{\partial\pi_{s,i}^2}=0$, 
$\forall s,i$. Therefore $\sigma^2$ is an harmonic function in $\DN$
%(i.e.$\sum_{i,s}\frac{\partial^2}{\partial\pi_{s,i}^2}\sigma^2=0$) 
which implies that the minima are on the boundary of $\DN$. This
holds for any subset of variables $\pi_{i,s}$ which therefore
implies that minima are located in the corners of the simplex,
i.e. Nash equilibria are in pure strategies ($G=1$). This, in its turn,
implies that Nash equilibria have $\sigma^2\approx H$ by Eq. \req{sigmaH}.

A detailed characterization of these Nash equilibria shall be given
elsewhere \cite{andemar}. Here we briefly mention that also in
this case Nash equilibria are exponentially many in $N$. 
This makes the analytic calculation a step more difficult than
the one we shall present later. 
A simplified approximate calculation (see appendix) gives the following 
lower bound\footnote{Strictly speaking, the meaning of this lower bound
is that the probability to observe $\sigma^2$ smaller than the
lower bound decreases exponentially with $N$}:
\be
\frac{\sigma^2}{N}\ge\left\{\begin{array}{lr}
\left[1-\frac{z(S)}{\sqrt{\alpha}}\right]^2 & \hbox{for $\alpha>
z(S)^2$}\\
0 & \hbox{for $\alpha\le z(S)^2$}
\end{array}\right.
\label{nash}
\ee
where $\alpha=P/N$ and $z(S)=\sqrt{2/\pi}S\int_{-\infty}^\infty 
\! dz\,e^{-z^2}z[1-{\rm erfc}(z)/2]^{S-1}$ is the expected value of 
the maximum among $S$ standard  random variable (for $S\gg 1$, 
$z(S)\simeq\sqrt{2\ln S}$).

\begin{figure}
\centerline{\psfig{file=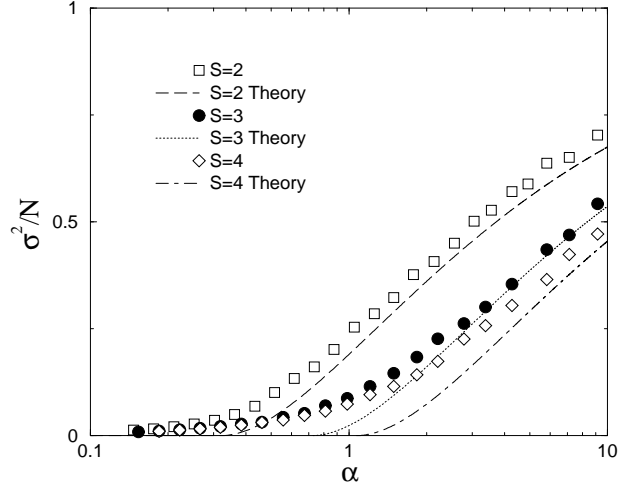,width=8cm}}
\caption{Global efficiency $\sigma^2/N$ as a function of $\alpha$
for $S=2,3$, and $4$ from numerical simulations with $P=128$
averaged over $100$ realizations of $\va_i$ (symbols) and from
the theoretical lower bound (lines).}
\label{fignash}
\end{figure}

Figure \ref{fignash} shows that the lower bound is already a good
approximation to the typical value of $\sigma^2$ in the Nash
equilibrium, specially for small values of $S$. 
Eq. \req{nash} implies that, for fixed $S$, 
$\sigma^2$ increases with $\alpha$, which is reasonable because
the complexity of information increases and the 
resources of agents is limited by $S$. For fixed $\alpha$,
Eq. \req{nash} suggests that $\sigma^2$ decreases with $S$.
So if agents are given more resources (larger $S$), they 
attain a better equilibrium. 
Both of these features are confirmed by numerical simulations 
(see fig. \ref{fignash}). 

It is worth to point out that the game specified by the payoffs 
of Eq. \req{ui2}, for
$N$ and $P$ very large, implies a fantastic computational
complexity. Deductive rational agents should be able to master a 
chain of logical deductions of formidable complexity in order
to derive their best response.
The efforts required by this strategic situation may well exceed 
the bounds of memory and computational capabilities
of any realistic agent or more simply the resources she is
likely to devote to the problem. Furthermore her assumption that
everybody else behaves as a rational deductive player becomes more
and more unrealistic as $N$ grows large. Finally, even with deductive
rational agents there would still be the problem of equilibrium
selection which, in this case, involves a huge number of possible
equilibria. 

\section{Repeated game: Learning and inductive rationality}

Deductive rationality, as suggested in refs. \cite{Arthur,CZ1,ZENews}, 
is unrealistic in such complex strategic situations\footnote{We do not
discuss learning and inductive rationality in simpler cases such as e.g.
$P=1$ and $\va_i=\A$ for all $i\in\N$. For a discussion of reinforcement 
learning in El Farol problem at this level see R. Franke (1999).} and it 
has to be replaced by inductive rationality. This amounts 
to assume that agents try to learn what their best choice is from their 
past performance. We henceforth focus on the repeated game in which
agents meet once and again to play the stage game of section \ref{gamedef}.
Different stage games are distinguished by the time label by $t\in\nit$.
For example, $s_i(t)$ denotes the strategy chosen by agent $i$ at time 
$t$ and $\mu(t)$ the information available at that time. 

\subsection{Exponential learning:}

It is generally accepted that agents follow more likely strategies which
have been more successful in the past, which is known\cite{roth,rust} 
as the ``law of effect''. There are several behavioral models implementing 
the law of effect (see e.g. \cite{roth,rust}). Here we assume that agents 
follow an {\em exponential learning} behavior: Each agent $i$ assigns {\em
scores} $U_{s,i}(t)$ to each of their strategies
$s=1,\ldots,S$ and she plays strategy $s$ with a probability
which depends exponentially on its score: 
\be
\pi_{s,i}(t)=\frac{e^{\Gamma_i U_{s,i}(t)}}
{\sum_{s'=1}^Se^{\Gamma_i U_{s',i}(t)}}
\label{explearn}
\ee
where $\Gamma_i >0$ is a numerical constant, which may differ 
for each agent $i\in\N$. This model for discrete choice -- called 
the {\em Logit model} -- has a long 
tradition in economics\footnote{This model is quite appealing since it 
satisfies the axiom of {\em independence from irrelevant alternatives} 
which states that the relative odds of choices $s$ and $s'$ does not 
depend on whether another choice $s''$ is possible or not.} \cite{logit} 
and some experimental support (see e.g. \cite{quantal}). 
The MG has been originally introduced
with $\Gamma_i=\infty$ \cite{CZ1,ZENews,Savit} and only recently it
has been generalized to $\Gamma_i<\infty$ \cite{cavagna2}. Note that
$\pi_{s,i}$ here is no more a mixed strategies -- which is the 
object of agents' strategic choice -- but rather it encodes a 
particular behavioral model. 

At time $t=0$ scores are set to some $U_{s,i}(0)$, which encodes 
{\em prior beliefs}:
e.g. $U_{s,i}(0)>U_{s',i}(0)$ means that agent $i$ considers
strategy $s$ {\em a priori} more likely to be successful than $s'$.

At later times $t>0$, agents update the scores of each 
of their strategies $s=1,\ldots,S$ in an additive way
\be
U_{s,i}(t+1)=U_{s,i}(t)+\Delta U_{s,i}^{\mu(t)}[t,s_i(t),s_{-i}(t)],
\label{updU}
\ee
where the {\em reinforcement} 
$\Delta U_{s,i}^{\mu(t)}[t,s_i(t),s_{-i}(t)]$ quantifies the 
``perceived'' success of each of their strategies $s=1,\ldots,S$ at 
time $t$. 
This generally depends both on the state $\mu(t)$ and on the strategies
$s_i(t)$ and $s_{-i}(t)$ played, at that time by agent $i$ and by
her opponents. 

\subsection{Naive and sophisticated agents}

What is the perceived success $\Delta U_{s,i}$ of a strategy $s$? 
The most natural way to quantify the success of a strategy is by 
the payoff it delivers to the agent if played. This  
suggests\footnote{The factor $1/P$ 
is introduced here and in the following equations for convenience. 
The reason will become clear in the next section.} that 
$\Delta U_{s,i}^{\mu}=u_i^{\mu}[s,s_{-i}]/P$ or 
\be
U_{s,i}(t+1)=U_{s,i}(t)+u_i^{\mu(t)}[s,s_{-i}(t)]/P.
\label{sophagents}
\ee
By this equation, however, one assumes that agent $i$ knows what payoff she 
would have got if she had 
played any strategy $s$, including those $s\ne s_i(t)$ which were not
used. In other words, agents must have {\em full 
information} on the effects (payoffs) of all of their strategies. 
Furthermore agents take into account the way in which
the aggregate quantity $A^{\mu(t)}$ would have changed if they had played 
strategy $s\neq s_i(t)$. Agents following Eq. \req{sophagents} are able of
sophisticated counter-factual thinking, and are henceforth called 
{\em sophisticated} agents\footnote{This term, as opposed to {\em naive}, 
is borrowed from ref. \cite{rust2}.}. 
It is worth observing that the score, with
the dynamics \req{sophagents}, acquires the meaning of cumulated payoff:
$U_{s,i}(t)$ is indeed the payoff agent $i$ would have received (divided
by $P$) if she had always played strategy $s$ against her opponents 
strategies $s_{-i}(t')$ for all $t'<t$.

In the minority game\footnote{and probably also in the El Farol problem, 
ref. \cite{Arthur} is not very clear on this point.} agents are {\em naive}
\cite{rust2}: {\em i)} they eventually have {\em partial information},
which means that they only know the payoff delivered by the 
strategy $s_i(t)$ which they actually played. {\em ii)} they behave as
if they were playing against an exogenous signal $A^{\mu(t)}$, rather 
than $N-1$ other agents. Naive agents neglect 
their impact on the aggregate $A^\mu$ and update scores {\em as if} $A^\mu$ 
had not changed if they had used a different strategy. More precisely
\be
\Delta U_{s,i}^{\mu(t)}[t,s_i(t),s_{-i}(t)]=-a_{s,i}^{\mu(t)}A^{\mu(t)}(t)/P.
\label{naiveupdMG}
\ee
Note that Eq. \req{sophagents} does not depend on the strategy
$s_i(t)$ which agent $i$ used, whereas Eq. \req{naiveupdMG} depends on
it because $A^{\mu(t)}(t)$ contains the action $a^{\mu(t)}_{s_i(t),i}$ 
which agent $i$ actually played.
Regarding the MG as a toy market, if the aggregate $A^{\mu(t)}(t)$ plays 
the role of price, Eq. \req{naiveupdMG} implies that agents behave 
as ``price takers'': They behave as if price did not depend on their
actions. By so doing, they simplify considerably the strategic complexity
of the context they face. Eq. \req{naiveupdMG} may be a 
closer approximation than Eq. \req{sophagents} to the behavior of real 
agents in complex strategic situations. 

One would naively expect that when $N\to\infty$ the 
difference between Eqs. \req{sophagents} and \req{naiveupdMG} 
is negligible. Indeed the relative impact of an agent on $A^\mu$ is 
negligible in that limit. Surprisingly we shall see that this is not so and 
a system of {\em naive} agents behave quite differently from 
{\em sophisticated} agents with full information. 
In order to study the effect of the impact of agent's choice on the
aggregate, we generalize Eq. \req{naiveupdMG} including a term 
$+\eta\delta_{s,s_i(t)}/P$. The dynamics of scores then reads
\be
U_{s,i}(t+1)=U_{s,i}(t)-a_{s,i}^{\mu(t)}A^{\mu(t)}(t)/P+
\eta\delta_{s,s_i(t)}/P.
%=-a^{\mu(t)}_{s,i}\sum_{j=1}^N a_{s_j(t),j}^{\mu(t)}+\eta\delta_{s,s_i(t)}.
\label{naiveupd}
\ee
The last term, which is absent ($\eta=0$) in the original definition
of the MG, models the tendency of agents to stick to the strategy they
are currently using. Indeed $\eta>0$ implies that agents reward the 
strategy they use $s=s_i(t)$ with respect to those they are not
currently using $s\ne s_i(t)$. By doing this, agents approximately
account for the impact of their actions on the global variable
$A^{\mu(t)}(t)$. As we shall see, this term has very deep consequences.

We shall consider these two cases -- of {\em partial information} with 
{\em naive} agents and of {\em full information} -- separately
below, and see that the collective behavior is remarkably
different. Before doing that, we shall first discuss the dynamics of
scores in the long run.

\section{Continuum time limit and the dynamics in the long run}

In this section, we shall first derive a continuum time dynamics for 
Eq. \req{updU} which captures the long run behavior of the system. 
Then we shall show that the collective behavior of agents, within this 
continuum time dynamics, admits a Lyapunov function, i.e. a functions which 
is minimized along the trajectories of the dynamics of the system. 
The dynamics therefore converges to the minima of this function.
This is a quite important step, since it allows to turn the study of the 
stationary state of the dynamical model 
into the study of the local minima of the Lyapunov function. Therefore one can
regard the Lyapunov function as the Hamiltonian of a system and resort to the
powerful tools of statistical mechanics in order to study the statistical
properties of its ground state (global minimum) 
and eventually of its meta-stable states (local minima). This
shall be the subject of the  next section.

In order to study the stationary state properties of the system, 
we need to consider the long time limit of the dynamics of scores,
Eq. \req{updU}. The key observation, in this respect, is that we
expect that $U_{s,i}(t)$ changes significantly and systematically
only over time-scales of order $\Delta t\sim P$. Indeed the score
of strategies depend on their performance on all the $P$ states
$\mu$. In order to capture the long time dynamics of scores, let
us set
\be
U_{s,i}(t)={\tilde U}_{s,i}(\tau),~~~\hbox{with}~~~\tau=\frac{t}{P}.
\ee
The dynamics in continuum time of ${\tilde U}_{s,i}$ is obtained 
iterating Eq. \req{updU} for $\Delta t=P d\tau$ time steps:
\be
\frac{{\tilde U}_{s,i}(\tau+d\tau)-{\tilde U}_{s,i}(\tau)}{d\tau}=
\frac{1}{P\,d\tau}
\sum_{t=P\tau}^{P(\tau+d\tau)-1} 
\Delta U_{s,i}^{\mu(t)}[t,s_i(t),s_{-i}(t)]
\label{dupsdt}
\ee
We can now take the thermodynamic limit $N,P\to\infty$ keeping
$d\tau$ finite. By the law of large numbers, the right hand side
converges almost surely to its average value (see later). Here 
$\Delta U_{s,i}^{\mu(t)}$ is a function of the random variables 
$\mu(t)$ and $\{s_j(t),~j\in\N\}$ which are chosen independently at each time 
step. If the stochastic fluctuations in ${\tilde U}_{s,i}$ are small
(see later) also the distribution $\pi_{s,i}(t)$ will be well 
behaved in the limit $P\to\infty$, specially if $\Gamma_i$ is
small\footnote{Interestingly, numerical simulations show that the 
continuum time approximation works generally also in the
limit $\Gamma_i\to\infty$.}. Indeed defining 
$\tilde\pi_{s,i}(\tau)=\pi_{s,i}(t=P\tau)$,  Eq. \req{explearn} becomes
\be
\tilde\pi_{s,i}(\tau)=\frac{e^{\Gamma_i {\tilde U}_{s,i}(\tau)}}
{\sum_{s'=1}^Se^{\Gamma_i {\tilde U}_{s',i}(\tau)}}
\label{explearncont}
\ee
which remains meaningful in the limit $P\to\infty$. Equivalently we
may say that the distribution $\vp_{i}$ remains approximately constant 
over $Pd\tau$ time steps.
Taking the continuum time limit $d\tau\to 0$ in Eq. \req{dupsdt}, we 
find
\be
\frac{d{\tilde U}_{s,i}}{d\tau}=\ovl{\avg{\Delta U_{s,i}}}
\label{conttime}
\ee
where averages are taken with respect to the distributions 
$\vp_{j}$ of strategies and $\rho^\mu$ of $\mu$. 

It is worth to point out that the order of the two limits --
first $P\to\infty$ and then $d\tau\to 0$ -- is quite important.
Indeed the infinitesimal time interval $d\tau$ corresponds
to a very large number $Pd\tau$ of time steps, which eventually
diverges. This implies that the characteristic time of the system
is proportional to $P$ time steps\footnote{In other words, $P$ 
repetition of the game is the analogous of a ``sweep'' of the system 
in a Montecarlo simulation.}.

The validity of the law of large number 
can be verified studying the fluctuations of $\Delta U_{s,i}^{\mu(t')}$ 
around its average $\ovl{\avg{\Delta U_{s,i}}}$. By routine use of
Tchebicev inequality, it is enough to show that 
\be
\frac{2}{(Pd\tau)^2}
\sum_{t,t'=P\tau}^{P(\tau+d\tau)-1}
%%%%%\sum_{t'=P\tau}^{P(\tau+d\tau)-1}
\Avg{\left(\Delta U_{s,i}^{\mu(t)}-
\ovl{\avg{\Delta U_{s,i}}}\right)\left(\Delta U_{s,i}^{\mu(t')}-
\ovl{\avg{\Delta U_{s,i}}}\right)}_{\mu,\{s_j\}}
\label{correl}
\ee
vanishes when $N,P\to\infty$. This is indeed the case because
the $\Delta U_{s,i}^{\mu(t)}$ depends on the random variables $\mu(t)$ 
and $\{s_j(t),~j\in\N\}$, which are drawn independently at each $t$
from their distributions $\rho^\mu$ and $\{\vp_j,~j\in\N\}$ respectively.
Only the terms with $t=t'$ or $\mu(t)=\mu(t')$ contribute to the 
average, whereas all other terms vanish because the average factorizes.
Only a fraction $\sim 1/P$ of the terms is non-vanishing, which implies 
that the expression \req{correl} is indeed of order $1/P$ and it 
vanishes as claimed.

We shall henceforth work with continuum time and drop the tilde over 
$U$ and $\pi$, in order to simplify notations.
Combining Eqs. \req{explearn} and \req{conttime}, 
and with a little algebra, one finds that $\pi_{s,i}$
satisfies the equation
\be
\frac{d\pi_{s,i}}{d\tau}=\gamma_i\pi_{s,i}\left[
\frac{d {U}_{s,i}}{d\tau}-
\vp_i\cdot\frac{d \vec {U}_{i}}{d\tau}\right]=\Gamma_i\pi_{s,i}
\left[\ovl{\Avg{\Delta U_{s,i}}}-
\vp_i\cdot\ovl{\Avg{\vec{\Delta U}_i}}\right].
\label{conttimep}
\ee

\subsection{Lyapunov function for naive agents}

The dynamics \req{naiveupd} of naive agents in continuum time
is easily derived combining Eqs.(\ref{naiveupd},\ref{conttimep}) and a
little algebra. It reads
\be
\frac{d\pi_{s,i}}{d\tau}=-\Gamma_i\pi_{s,i}
\left\{\sum_{j\in\N}
\left[\ovl{a_{s,i}\avg{a_j}}-
\ovl{\avg{a_i}\avg{a_j}}\right]
-\eta(\pi_{s,i}-|\pi_i|^2)           %\vp_i\cdot\vp_i)
\right\}
\label{naivedyn}
\ee
with the shorthand $\avg{a_j^\mu}=\vp_j\cdot\va_j^\mu$.
This is different from RD (Eq. \ref{RD}). So $\vp_i$ does not converge, 
in general, to a Nash equilibrium.  Rather one can show that 
\be
H_\eta=H-\eta\sum_{i\in\N}|\pi_i|^2%\vp_i\cdot\vp_i
\label{Heta}
\ee
is a Lyapunov function of this dynamics. Indeed, observing that
\[
\frac{\partial H_\eta}{\partial \pi_{s,i}}=-2\frac{dU_{s,i}}{d\tau}
\]
and using Eq. \req{conttimep}, one finds that
\be
\frac{dH_\eta}{d\tau}=\sum_{i\in \N}\frac{\partial H_\eta}{\partial \vp_i}
\cdot\frac{d\vp_i}{d\tau}=
-2\sum_{i\in\N}\Gamma_i\sum_{s=1}^S\pi_{s,i}
\left(\frac{dU_{s,i}}{d\tau}-\vp_i\cdot \frac{d\vec U_{i}}{d\tau}\right)^2<0.
\label{lyapHeta}
\ee
The dynamics converges therefore to the minima of $H_\eta$. 

This equation implies that in the stationary state ($dH_\eta/d\tau=0$) 
each of the
strategies played by agent $i$ -- those with $\pi_{s,i}>0$ -- has the same 
perceived success $\frac{dU_{s,i}}{d\tau}$ in the long run. 
Note also that for $\eta=1$, by Eq. \req{sigmaH}, $H_1\simeq \sigma^2-N$
which implies that for $\eta=1$ the stationary state is close to a Nash
equilibrium.

We shall come back to the statistical characterization of the stationary state
for naive agents in the next section. It is worth to stress, at this point that
this result holds for any realization of $a_{s,i}^\mu$. It actually holds for 
much more general models \cite{universality}. 

\subsection{Agents with full information}

It is known \cite{rust} that exponential learning
with full information, for a single agent playing against a
stationary stochastic process, converges to rational
expectations. If we can regard the opponents of $i$ as a
stationary process, this implies that $i^{\rm th}$ strategy 
converges to the best response. If this happens for all players
the system converges to a Nash equilibrium. This is indeed what
numerical experiments show (see figure \ref{fignash}).

We can recover this result within the continuum time limit. Indeed, after
some algebra one finds that Eq. \req{conttimep} becomes, in this case 
\be
\frac{d\pi_{s,i}}{d\tau}=-\Gamma_i\pi_{s,i}\sum_{j\in\N,\,j\ne i}
\left[
\ovl{a_{s,i}\avg{a_j}}
-\ovl{\avg{a_i}\avg{a_j}}
\right].
\label{RD1}
\ee
Apart from the factor $\Gamma_i$, this coincides with the 
RD of Eq. \req{RD}. Again $\sigma^2$ is minimized along the
trajectories of Eq. \req{RD1}: it is easy to check that the
time derivative of $\sigma^2$ is given by Eq. \req{lyapsig}
with an extra factor $\Gamma_i$ inside the sum on $i\in\N$.
We therefore conclude that {\em with exponential learning
and full information agents coordinate on a Nash equilibrium}.
Each agent plays, in the long run a pure strategy, i.e. $G=1$. 
The Nash equilibrium to which agents 
converge depends on the initial conditions $U_{s,i}(0)$, 
i.e. on prior beliefs: Different initial conditions select different
Nash equilibria.

\section{Statistical mechanics of naive agents}

As we have shown, the stationary state of the system is described by
the ground state of the Hamiltonian $H_\eta$. This can be analyzed using the 
tools of statistical physics and in particular, the replica method which allows
us to deal with quenched disorder (i.e. agents' heterogeneity). The details of
the calculation are described in the appendix in some detail. Here we shall
just describe the results. We shall consider separately the results for
$\eta=0$, i.e. for the original MG, for which we can derive exact results
within a relatively simple calculation. The case $\eta>0$ requires more complex
calculations which shall be the subject of a forthcoming paper \cite{andemar}.
Here, the qualitative understanding for $\eta>0$ provided by the present 
approach will be supplemented by numerical simulations. 

\subsection{$\eta=0$: The minority game}

It is easy to see that for $\eta=0$, the Hamiltonian $H_0\equiv H$ is a
non-negative quadratic form of the variables $\pi_{s,i}$ and therefore it
attains its minimum  on a connected subset ${\cal M}\in\DN$\footnote{Note that,
on the contrary, $\sigma^2$ is not positive definite and
it attains its minima, the Nash equilibria, on a non-connected
subset of $\DN$.}.
We therefore conclude that {\em the long run dynamics of 
this system is described by the minimum of $H$}. Loosely speaking, in view of 
our definition of $H$, we may say that naive agents, in the minority game, 
minimize the ``information content'' of the market output $A^{\mu(t)}(t)$.

A complete statistical characterization of the minima of $H$ in the 
limit $N\to\infty$ with $P/N=\alpha$ finite and $\varrho^\mu=1/P$, can
be obtained from the {\em replica method} a tool of 
statistical mechanics devised to deal with disordered systems. An
account of this method is given in the appendix together with
technical details on the calculation. Here we only discuss the 
results and their interpretation,
We distinguish two regimes separated by a {\em phase transition} which
occurs as $\alpha\to\alpha_c(S)\cong S/2-0.6626\ldots$. 

\subsubsection{Asymmetric phase: $\alpha>\alpha_c$}

For $\alpha>\alpha_c$ we find an {\em asymmetric phase}. Indeed $H>0$ which
means that $\avg{A^\mu}\ne 0$ at least for some $\mu\in\P$. The symmetry 
between the two actions in $\A$ is broken and a {\em best}
strategy $a^\mu_{\rm best}=-\sign\avg{A^\mu}$ arises in $\AP$. 
An $N+1^{\rm st}$ agent
who joined the game with this strategy would receive a payoff
$\ovl{|A|}-\ovl{a^2_{\rm best}}=
\ovl{|A|}-1$\footnote{The term $-\ovl{a^2_{\rm best}}$ is the 
``market'' impact caused by the new agents. It arises
because if the strategy $a^\mu_{\rm best}$ where 
actually played by the $N+1^{\rm th}$ agent, that would also modify 
$A^\mu\to A^\mu+a^\mu_{\rm best}$. 
Ref. \cite{prodspec} discusses in greater detail 
these issues.}. 

The set ${\cal M}$ where $H$ attains its minimum consists of a single 
point, so that, for any initial conditions, the dynamics converges 
to the same final state in the long run. In other words prior 
beliefs of agents about their strategies are irrelevant 
in the long run. 

The asymmetry $H$ decreases with decreasing $\alpha$, which means
increasing the number $N$ of agents at fixed $P$. Naively speaking 
the asymmetry in $\avg{A^\mu}$ is exploited by the adaptive behavior
of agents who then reduce it. Indeed agents are more and more
selective in their choice, as shown by the fact that $G$ increases
as $\alpha$ decreases %(see figure \ref{figG}) 
and the effective 
number of strategies used $1/G$ decreases.
At the same time, as $\alpha$ 
decreases, the equilibrium becomes more and more 
``fragile'' in the sense that its {\em susceptibility} to a
generic perturbation increases (see the Appendix and ref. \cite{MPV}). 

\subsubsection{Phase transition and symmetric phase: $\alpha<\alpha_c$}

As $\alpha\to\alpha_c$ the asymmetry vanishes $H\to 0$ and 
the response of the system to a generic perturbation diverges.
This signals a phase transition to the symmetric phase $\alpha<\alpha_c$ 
where $H=0$ and any perturbation can change dramatically the 
equilibrium. The set ${\cal M}$ where $H$ attains its minimum
is no more a single point, but rather an hyper-plane of a dimension
which increases as $\alpha$ decreases. Any point in ${\cal M}$ is
an equilibrium of Eq. \req{naivedyn} and any displacement along
this set can occur freely. In particular, with different initial 
conditions, the system reaches different points of ${\cal M}$.
The dynamics \req{naivedyn} indeed converges to the 
``closest'' point on ${\cal M}$ which is on its trajectory.
In other words, prior beliefs $U_{s,i}(0)$ are relevant for 
$\alpha<\alpha_c$: With different $U_{s,i}(0)$ the system
reaches different equilibria.

\begin{figure}
\centerline{\psfig{file=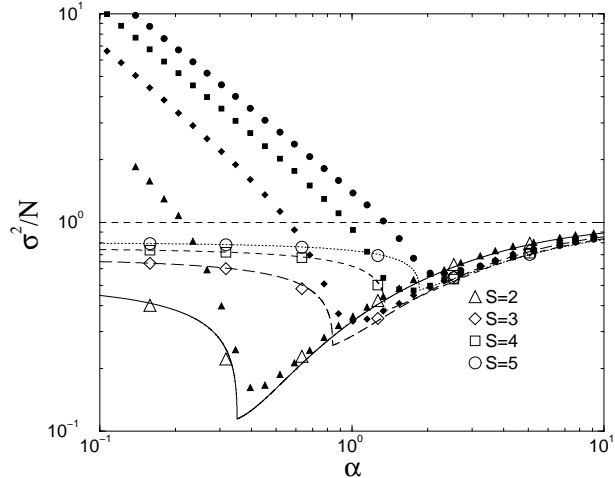,width=8cm}}
\caption{Global efficiency $\sigma^2/N$ as a function of $\alpha$
for $S=2,3,4$ and $5$ from numerical simulations of the minority game
with $N=101$ agents and $\Gamma_i=\infty$ 
(averages were taken after $100P$ time steps), 
averaged over $100$ realizations of $\va_i$ 
(small full symbols), from numerical minimization of $H$ (large open 
symbols) and from the theoretical calculation (lines) with 
$U_{s,i}(0)=0$.}
\label{figsigma}
\end{figure}

\subsubsection{Anti-persistence in the symmetric phase}

For $\alpha<\alpha_c$ the system is dynamically degenerate: any
displacement on the set ${\cal M}$ can occur freely. In particular
stochastic fluctuations can induce a motion in ${\cal M}$. 
This is what happens for $\Gamma_i\gg 1$ where numerical simulations
show the presence of ``crowd effects'' \cite{Savit,CZ2,johnscrowd}.
This effect manifests in an increases of $\sigma^2/N$ as $\alpha$ decreases,
which is much faster than what predicted by our theory (see full 
symbols in fig. \ref{figsigma}). This behavior can 
be traced back to a dynamical anti-persistence \cite{CM} resulting
from the fact that agents, neglecting their impact on $A(t)$, 
over-estimate the performance of the strategies they do not play
and they keep switching from one to the other. Each time a 
particular state $\mu$ shows up, agents tend to do the 
opposite action of what they did the last time they saw the same 
state $\mu$. 
Therefore the period of this dynamics is of $2P$ time steps\footnote{The fact 
that agents do not realize that they have an impact on the
aggregate in this case is probably unrealistic. From their point of 
view, the same strategy which had a good score of
performance when they were not using it, starts performing badly
as soon as they use it.  
This could either be considered a manifestation of {\em Murphy's law}
or the fact that agents rationality is bounded below the level of
common sense.} \cite{CM}. 

The analytic approach to this effects requires the study of the 
dynamical solutions of Eq. \req{naiveupd}, which go beyond the aims of
the present work. We suspect that one should refine our continuum 
time approach, including eventually the second order time derivative
and the effects of fluctuations to some extent. Indeed a periodic 
motion is usually related to the inertial term ($d^2U/d\tau^2$) of
the dynamic equation.

The periodic behavior persists as long as $1/\Gamma_i$ is much smaller than
the amplitude of the oscillations of $U_{s,i}(t)$.
As $\Gamma_i$ decreases, Eq. \req{explearn} finally smoothes 
the oscillations in agents choices and the anti-persistent behavior 
disappears, as indeed observed in \cite{cavagna2}. 

\subsubsection{Global efficiency and Tragedy of Commons}

As far as global efficiency is concerned, we find that $\sigma^2/N$ 
increases with $\alpha$ towards the random agent limit as
$1-\sigma^2/N\sim 1/\alpha$. 
This is shown in figure \ref{figsigma}, which also shows that 
numerical simulations for finite populations fully confirm our 
theoretical results. For fixed $\alpha$, as $S$ increases $\sigma^2/N$
first decreases moderately as long as $\alpha >\alpha_c(S)$. 
Then the system enters the symmetric phase ($\alpha<\alpha_c$) 
because $\alpha_c(S)$ increases, and
$\sigma^2/N$ increases with $S$ towards the random agent limit
as $1-\sigma^2/N\sim 1/S$. We then conclude that allowing agents
to have more strategies does not increase global efficiency as
in the Nash equilibrium. Rather it pushes the system in the 
symmetric phase where $\sigma^2$ converges to the random agent
limit as $S\to\infty$. 

As long as the system is in the asymmetric phase, agents
have incentives to consider more than $S$ strategies, because that
allows them to detect better the asymmetry. However if every agent
enlarges the set of strategies she considers, i.e. if $S$ increases,
the system enters in the symmetric phase. Then global efficiency 
starts to decline. This behavior is reminiscent of the {\em Tragedy
of the Commons} \cite{hardin}: a situation where 
individual utility maximization by all agents leads to
over-exploitation of common resources and poor payoffs.

\subsection{Rewarding the played strategy: $\eta>0$}

We expect that with $\eta=1$ agents behave almost optimally, in the
sense that they converge to a stationary state which is close to a
Nash equilibrium. Given the difference in the collective behavior of 
agents in the two cases -- which may be appreciated comparing figure
\ref{figsigma} for $\eta=0$ with figure \ref{fignash} for $\eta=1$ -- it is
natural to ask what happens when $\eta$ changes continuously from
$0$ to $1$.

\begin{figure}
\centerline{\psfig{file=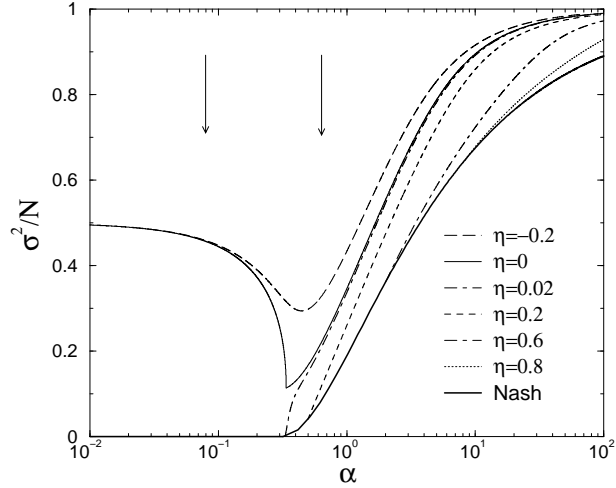,width=8cm}}
\caption{Theoretical estimate of global efficiency $\sigma^2/N$ as a function 
of $\alpha$ for $S=2$ and several values of $\eta$ within the {\em replica 
symmetric ansatz}.}
\label{figeta_RS}
\end{figure}

Figure \ref{figeta_RS} shows the analytical prediction for the dependence
on $\eta$ of $\sigma^2/N$ for $S=2$. These are based on the {\em replica 
symmetric ansatz} which is only valid for $\alpha>\alpha_{\rm RSB}(\eta)$,
where $\alpha_{\rm RSB}(\eta)$ marks a {\em replica symmetry breaking} phase
transition, which will be discussed elsewhere in detail\cite{andemar}.
Here we just mention that $\alpha_{\rm RSB}(\eta)=0$ for $\eta<0$,
$\alpha_{\rm RSB}(0)=\alpha_c$ and 
$\alpha_{\rm RSB}(\eta)\ge 1-1/\sqrt{\pi\alpha}$ (for $S=2$ and) $\eta>0$.
For $\alpha<\alpha_{\rm RSB}(\eta)$ the analytical results derived in the 
appendix provides an approximate description of the behavior of the system
which is however sufficient to appreciate the relevant features. 

\begin{figure}
\centerline{\psfig{file=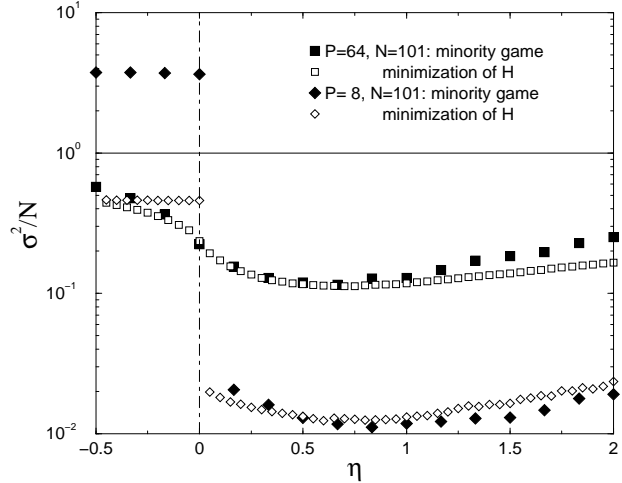,width=8cm}}
\caption{$\sigma^2/N$ as a function of $\eta$ for $S=2$ and
$\alpha\simeq 0.079<\alpha_c\simeq 0.3374$ and $\alpha\simeq 0.63>\alpha_c$.
Results both of numerical simulations of the minority game and of 
the numerical minimization of $H_\eta$ are shown.}
\label{figsigvseta}
\end{figure}

The most striking consequence of the result in fig. \ref{figeta_RS} is that
the behavior of $\sigma^2/N$ is quite different for $\alpha>\alpha_c$ and
for $\alpha<\alpha_c$. Indeed for large $\alpha$, 
$\sigma^2/N$ changes continuously
with $\eta$ whereas $\sigma^2/N$ drops discontinuously to zero as $\eta\to 0$ 
for small $\alpha$. This feature is reproduced in figure \ref{figsigvseta}
for two characteristic values of $\alpha$ also shown as arrows
in fig. \ref{figeta_RS}. We show both the behavior derived from the numerical 
minimization of $H_\eta$ and the behavior of the modified minority game with
rewarding. Numerical results agree quite well in an intermediate range of
values of $\eta$ whereas for $\eta>0.5$ or $\eta<-0.2$ some discrepancy --
which we believe is due to finite size effects -- is found. 
This effect can be even more spectacular when anti-persistence effects occur.
Indeed the jump of $\sigma^2/N$ at $\eta=0$ can be of several orders of 
magnitude! 

The origin of this behavior lies in the dynamic degeneracy of the system
for $\alpha<\alpha_c$ and $\eta=0$. Even an infinitesimal change in $\eta$
can dramatically alter the nature of the minima of $H_\eta$: for
negative $\eta$ there is only one minimum which becomes shallower and 
shallower as $\eta\to 0^-$. At $\eta=0$ the minimum is always unique but
it is no more point-like. Rather it is a connected set ${\cal M}$. 
An infinitesimal positive value of $\eta$ is enough to lift this 
degeneracy and select only some extreme points of ${\cal M}$ as the minima
of $H_\eta$. The set of minima becomes suddenly disconnected.
At fixed $\alpha<\alpha_c$, varying $\eta$ across the transition 
$H_\eta$ changes continuously -- with a discontinuity in its first
derivative -- whereas $G$ and hence $\sigma^2/N$ change discontinuously 
with a jump.

The potential implications of this result are quite striking: {\em rewarding
the strategy played more than those which have not been played by a small
amount is always advantageous, both individually} (see below) {\em and 
globally}. In particular,
{\em an infinitesimal reward is sufficient to avoid crowd effects when
$\alpha$ is small and to reduce the fluctuations by a finite amount}.

\section{From the agent's viewpoint}
\label{agentview}

Let us consider the behavior of agents in some detail (see also ref. 
\cite{prodspec} for a related discussion). Our goal is to 
show that individual payoffs increase when $\eta$ increases in
$[0,1]$. This means that rewarding the strategy ``in vivo'', that which
has actually been played, is convenient for each agent.
We focus on one agent, say $i$, and assume that others play strategies
$s_{-i}(t)$ according to some stationary probability distribution 
$\vp_{-i}$. If $\mu(t)$ is drawn randomly from $\rho^\mu$ we can consider
$A_{-i}^{\mu(t)}(t)=\sum_{j\ne i} a_{s_j(t),j}^\mu(t)$ as a stationary 
process. Since we deal with one agent, we shall drop the subscript $i$
in this section. Also we first focus on the case $\eta<1$ and only after
discuss the case $\eta>1$.
In the long run the perceived performance of strategy $s$ is
\bea
\Avg{\Delta U_s}&=&-\ovl{a_s\Avg{A_{-i}}}-\vp\cdot\ovl{\va a_s}+\eta\pi_s
\nonumber\\
&\cong &-\ovl{a_s\Avg{A_{-i}}}-(1-\eta)\pi_s
\label{eqxx1}
\eea
where the approximation in Eq. \req{eqxx1} holds for $P\gg 1$ since
$\ovl{a_{s'}a_s}\sim 1/\sqrt{P}$ for $s'\ne s$. Because of Eq. \req{conttimep},
strategies can either {\em i)} have $\pi_s>0$ and $\Avg{\Delta U_s}=v$ 
independent of $s$ or {\em ii)} have $\pi_s=0$ and $\Avg{\Delta U_s}<v$.
This can be understood by a rather simple argument: 
Imagine that strategy $1$ has $\Avg{\Delta U_1}>v$. Then by the very learning 
dynamics, the agent shall use strategy $1$ more frequently than others
and hence $\pi_1$ shall increase. 
Because of the last term in Eq. \req{eqxx1} that will decrease
the perceived performance $\Avg{\Delta U_1}$ of that strategy. On the other
hand, if $\Avg{\Delta U_1}<v$ the agent shall use it less frequently, hence
its $\Avg{\Delta U_1}$ shall increase. If $\Avg{\Delta U_1}<v$ even when 
$\pi_1\to 0$ then the agents will never play 
that strategy, i.e. $\pi_1=0$. 

Let $n\le S$ be the number of strategies with $\pi_s>0$ and let these be
labeled by $s=1,\ldots,n$, whereas $\pi_k=0$ for $k>n$. Taking the sum
of Eq. \req{eqxx1} on $s=1,\ldots,n$ we find
\[
v=-\frac{1}{n}\sum_{s=1}^n \ovl{a_s\Avg{A_{-i}}}-\frac{1-\eta}{n}
\]
where $n$ is fixed by the condition $v>-\ovl{a_k\Avg{A_{-i}}}$ for all
$k>n$. Clearly $-\ovl{a_s\Avg{A_{-i}}}>v>-\ovl{a_k\Avg{A_{-i}}}$ for any
$s\le n$ and $k>n$ hence the $n$ strategies which the agent uses are the 
$n$ more efficient ones. Then Eq. \req{eqxx1} becomes 
\[
\pi_s=\frac{1}{n}+\frac{1}{1-\eta}
\left(\frac{1}{n}\sum_{s'=1}^n \ovl{a_{s'}\Avg{A_{-i}}}-
\ovl{a_s\Avg{A_{-i}}}\right).
\] 
Note that strategies with a larger 
$-\ovl{a_s\Avg{A_{-i}}}$ are played more frequently.
The average payoff $\ovl{u}=-\vp\cdot\ovl{\va\Avg{A_{-i}}}-1$ 
delivered by a learning behavior with parameter $\eta$ is 
\be
\ovl{u}=-\frac{1}{n}\sum_{s=1}^n \ovl{a_s\Avg{A_{-i}}}+
\frac{1}{1-\eta}\sum_{s=1}^n\left(\ovl{a_s\Avg{A_{-i}}}
-\frac{1}{n}\sum_{s'=1}^n \ovl{a_{s'}\Avg{A_{-i}}}\right)^2-1
\ee
which is an increasing function of $\eta$ for $\eta<1$. Indeed at fixed $n$,
this is trivially true. With some more algebra, it is easy to
check that $n$ is a non-increasing function of $\eta$ and that
$\ovl{u}$ increases as $n$ decreases. This means that {\em for
$\eta<1$ average payoffs are non-decreasing functions of $\eta$}
as claimed.

When $\eta\to 1$ the only possible solution is that with $n=1$ which
means that the agent plays her best response to $A_{-i}$. For $\eta>1$
the agent over-weights the performance of her strategies. As a result
she sticks to only one of her strategies, i.e. $n=1$, but that need not
be her best one. Without entering in too many details, let us only mention
that for $\eta>1$ the agent plays always one strategy which is dynamically
selected by initial conditions and stochastic fluctuations.

\section{Exogenous vs endogenous information}
\label{endoexo}

In the El Farol problem and in the MG the state $\mu(t)$ is determined by
the outcome of past games. In other words $\mu(t)$ is an {\em endogenous} 
information which encodes information on the game itself: Agents record 
which has been the winning action in the last $M=\log_2 P$ games and store
this information in the binary representation of the integer $\mu$.
This means that $\mu$ is updated at each time as:
\be
\mu(t+1)={\rm mod}\left(2\mu(t)+\frac{1+\sign
A(t)}{2},P\right),~~~~A(t)=\sum_{i\in\N}
a_{s_i(t),i}^{\mu(t)},
\label{updmu}
\ee
Note that $A(t)$ depends on time both through $\mu(t)$ and through
the choice $s_i(t)$ of each agent $i\in\N$ at time $t$.
Eq. \req{updmu} implies that the dynamics of $\mu(t)$ 
is defined by the collective behavior of the game itself. 
Still payoffs do not depend on $\mu(t)$ which is a sun-spot.
However agents can coordinate in such a way that some state 
$\mu$ -- i.e. some pattern in the time series of $A(t)$ -- can
occur more frequently than some other and eventually some can 
never occur. As far as the collective behavior in the stationary state
is concerned, the only relevant information of this dynamics is the 
stationary state distribution $\varrho^\mu$ of the process 
Eq. \req{updmu}. In the long run, the distribution 
$\varrho^\mu$ is determined by the collective behavior of agents through 
$A(t)$. Technically, the problem of computing $\varrho^\mu$ 
is related to the diffusion of a 
particle on the directed graph defined by Eq. \req{updmu}, where
each note $\mu\in\P$ has two outgoing links to nodes ${\rm mod}(2\mu,P)+1$
and ${\rm mod}(2\mu+1,P)+1$ (and two incoming links). Stochastic 
processes on this graph, known as De Bruijn graph, are called shift register 
sequences \cite{debruijn} in computer science.

This differs from the setting we have discussed so far, in which $\mu(t)$ is 
independently drawn at each time $t$ with $\varrho^\mu=1/P$.
We may call $\mu$ {\em exogenous information} in this case since it can be 
considered to encode
information about an external system, eventually the environment
where agents live. This version of the MG has been first introduced 
by Cavagna \cite{cavagna}. He found that in 
numerical simulations the collective behavior with exogenous information 
differs only weakly from that under endogenous information. 
Having already discussed the results for the exogenous case, let us now
consider how these change under endogenous information.

\subsection{Naive agents with $\eta\le 0$ and endogenous information}

With endogenous information the system behaves qualitatively 
in the same way, as first observed in ref. \cite{cavagna} by
numerical simulations for $\eta=0$. This is because the stationary
state distribution $\varrho^\mu$ of the process $\mu(t)$ -- which
is induced by the dynamics of agents through Eq. \req{updmu} -- 
is almost uniform on $\P$. Actually $\varrho^\mu=1/P$ for 
$\alpha\le\alpha_c$ because of the symmetry of $A^\mu$. 

\begin{figure}
\centerline{\psfig{file=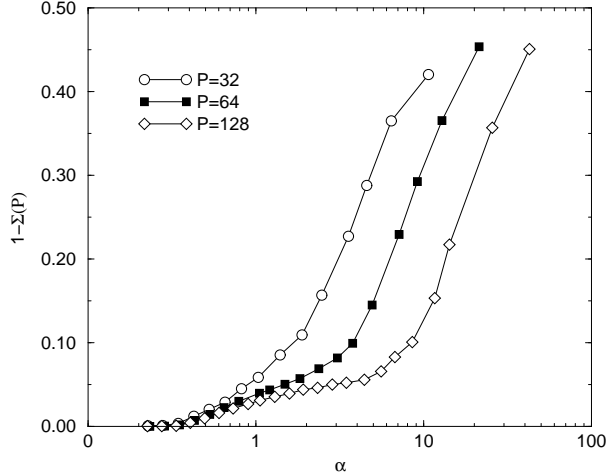,width=8cm}}
\caption{Deviation of the distribution $\varrho^\mu$ from the uniform
one from numerical simulations with $\eta=0$.}
\label{figentropy}
\end{figure}

In order to measure the deviation of $\varrho^\mu$ 
from the uniform distribution $\varrho^\mu_{\rm unif}=1/P$, we 
compute the entropy $\Sigma(P)=-\sum_{\mu\in \P}\varrho^\mu\log_P
\varrho^\mu$. With the choice of base $P$ for the logarithm
$\Sigma(P)=1$ for $\varrho^\mu=1/P$ so that 
$1-\Sigma(P)$ is a reasonable measure of the deviation of 
$\varrho^\mu$ from a uniform distribution. 
In figure \ref{figentropy}, $1-\Sigma(P)$ is plotted 
for several values of $P$ as a function of $\alpha$. While for
$\alpha<\alpha_c$ we find $\Sigma(P)=1$ to a great accuracy, 
for $\alpha>\alpha_c$ numerical results suggest that $\Sigma(P)\to 
1$ as $P=\alpha N\to\infty$. On this basis, we conclude as in ref. 
\cite{cavagna}, that the MG with endogenous information gives
the same results as the MG with exogenous information. 
By a detailed study of the dynamics of the process $\mu(t)$
one can actually give a deeper theoretical foundation to this 
conclusion and derive analytically this result \cite{elsewhere}.
We conjecture that, as long as agents choice remains stochastic ($G<1$),
as for $\eta\le 0$, the dynamics in $\mu(t)$ is ergodic in $\P$. 
Note indeed that, for any $\mu$, $A(t)$ has stochastic fluctuations 
around its average value $\avg{A^\mu}$ which are of the same order
of magnitude of the average itself.

\subsection{$\eta>0$ and agents with full endogenous information}

A qualitatively different situation arises when $\eta>0$ and in particular
when agents have full information. We shall mainly discuss the latter case
which corresponds to $\eta=1$ and then discuss briefly the generic $\eta>0$
case. The key observation is that agents for $\eta=1$ play pure strategies
($G=1$). This means that agents behave the same whenever $\mu(t)=\mu$
and accordingly $A(t)$ shall always take the same value $A^\mu$ 
each time $\mu(t)=\mu$. This implies that the dynamics of Eq.
\req{updmu} for $\mu(t)$ becomes deterministic. More precisely it
locks into a periodic orbit $\mu(t+T)=\mu(t)$ with some
period $T$. Only the values of $\mu$ into this orbit shall
occur in the long run, whereas all other values of $\mu$ shall
never occur. This means that agents strongly influence
the time series of $\mu(t)$ and hence of the aggregate $A(t)$. Most remarkably,
in doing so, they achieve a much better coordination with respect
to the exogenous information case because they reduce the
parameter $\alpha=P/N$ to $\tilde \alpha=T/N$\footnote{Note
that the values of $\mu$ which occur in the long run are sampled
uniformly.}. Numerical studies show that $T\propto\sqrt{P}$ 
so that, in the limit $N\to\infty$ with 
$P/N=\alpha$ finite, $\tilde\alpha\propto 1/\sqrt{N}$ and also
$\sigma^2/N$ takes a very small value\footnote{The result 
$T\propto\sqrt{P}$ is what one would obtain on a random directed 
graph with $P$ vertex each with two outgoing links. This is a reasonable
approximation because the dynamics locks into periodic loops of 
$T\propto\sqrt{P}$ vertices where the peculiar structure of De Bruijn
graphs (see Eq. \req{updmu}) does not play a significant role.}.

The same behavior shall be expected for all $\eta>0$ such that $G=1$.
Therefore we expect that for each $\alpha$ there shall be a particular
value $\eta_{\rm EB}(\alpha)$ beyond which ergodicity of the dynamics 
of $\mu(t)$ in $\P$ breaks down. For values of $\eta>\eta_{\rm EB}(\alpha)$ 
we expect the dynamics of $\mu$ to lock into periodic orbits causing 
a reduction of $\sigma^2$ similar to that discussed above.

\section{Discussion}

There is a 
growing literature on learning which addresses the issue of
which learning procedure (modeling inductive rationality) 
may eventually lead to deductive rational outcomes\cite{learn,rust}.
The choice made in the El Farol bar problem and in the 
MG -- which is exponential learning (Eq. \ref{explearn}) 
eventually with $\Gamma_i\to\infty$ -- is one of these as we have
shown. What leads to equilibria different from Nash equilibria
is the fact that agents {\em i)} have not full information
on the effects of their strategies and {\em ii)} that they 
neglect or do not account properly for their impact on the aggregate 
in the evaluation of their strategies. 

In particular, for $\eta=0$, this new equilibrium, which we call 
{\em naive agents equilibrium} (NAE) differs substantially from a Nash 
equilibrium (NE), because:
\begin{enumerate}
\item In a NE global efficiency always increases as the number 
$N$ of agents increases (with $P$ fixed), whereas in the NAE
it only decreases as far as $N< P/\alpha_c$ and then it 
increases in a way which depends on initial conditions (prior
beliefs) and on the parameters $\Gamma_i$.
\item There are (exponentially) many {\em disconnected} Nash 
equilibria which are selected by initial conditions, i.e. by prior
beliefs. For 
$\alpha>\alpha_c$ there is a unique NAE and, for all initial 
conditions, the system converges to it. For $\alpha<\alpha_c$ 
there is a continuum of NAE, but they are all connected.
\item Global efficiency ($\sigma^2$), for fixed $\alpha$, 
always decreases as agents 
resources ($S$) increase, and it eventually converges to perfect 
optimization for $S\to\infty$. In the NAE efficiency only
mildly improves with $S$ in the asymmetric phase. But increasing $S$ 
also  increases $\alpha_c(S)$ and when $\alpha_c(S)>\alpha$ the
system enters into the symmetric phase where $\sigma^2$ increases
with $S$ towards the random agent limit (this occurs for $\Gamma_i$ 
small \cite{cavagna2}). 
\item In the NE agents play {\em pure} strategies, i.e. $\vp_i$
is a singleton $\forall i\in\N$. Indeed, for fixed opponent 
strategies $s_{-i}$, each agent typically\footnote{Typically here
means that we disregard unlikely realizations where two pure strategies
happen to yield the same payoffs.} has a pure strategy $s_i$ -- 
the best response -- which is superior to others.  In the NAE, 
agents mix this strategy with others because, 
neglecting their impact on $A(t)$, they over-estimate the 
performance of the strategies they do not play.
Playing a pure strategy reduces its perceived performance and this is
why agents mix strategies in the NAE (see also ref. \cite{prodspec}). 
The probability $\pi_{s,i}$ in the NAE
is such that the perceived performance of all strategies which
are played ($\pi_{s,i}>0$) is the same.
\item A consequence of the previous point is that, the origin of 
information is quite important for inductive agents with full
information, while it is irrelevant in the NAE \cite{cavagna}.
Inductive agents with full information lead, under {\em endogenous} 
information, to a deterministic dynamics of $\mu(t)$ and only
a small subset of informations $\mu$ is ever visited. This in
turn leads to a much more efficient coordination. 
Naive agents with endogenous information, on the other hand,
induce a dynamics on $\mu(t)$ which is ``ergodic'', i.e. which
visits each information $\mu$ with nearly the same frequency
$\varrho^\mu$. Therefore the collective behavior is the same as 
that of the NAE with exogenous information.
\end{enumerate}

For intermediate values of $\eta$ the collective behavior of agents 
interpolates between these two situations in a continuous way for
$\alpha>\alpha_c$ or in a discontinuous way for $\alpha<\alpha_c$.

We expect that rewarding the strategy which is
played with respect to those which are not played should be advantageous
both individually and globally in more general situations where agents
interact through a global variable {\em via} a minority-like mechanism.
Indeed one can show that the qualitative picture we have described 
remains the same if we allow for heterogeneity of various sorts such 
as allowing for a dependence on $i$ of $S$ and $\eta$,
or changing the distribution in Eq. \req{probasim} to a 
generic $P_i(a)$ for $a\in\rit$ \cite{universality} (see the appendix).

Our results clearly allow for several extensions, as those
of ref. \cite{prodspec}. It also suggests a theoretical approach to 
the El Farol problem \cite{Arthur}\footnote{The key problem lies in the
parametrization of forecasting rules: agents in the El Farol 
problem consider the record of past attendance to the bar
-- i.e. $A(t')$ for $t'<t$ -- whereas in the minority 
game agents only consider the record of the sign of $A(t')$.
Focusing on the last $M$ games, there can be $N^M$ possible records 
in the El Farol problem, instead of $2^M$. This causes no problem
in principle since one can take $P=N^M$. In practice however forecasting
rules have to be ``reasonable'':
For example a randomly drawn rule can easily predict a different outcome
if the attendance of one of the past weeks just changes by one unit.
Some sort of continuity in strategies should be introduced so 
that ``similar'' histories (=information $\mu$) lead to similar
forecasts. Even though it is not clear how to translate this
requirement mathematically 
(one way could be that followed in ref. \cite{mercatino}) it
is clear that it suggests that the number of relevant 
informations $P$ is much less than $N^M$.}. 
We expect that the distinction between inductive agents with 
full information and naive agents to be of key importance
also in the El Farol problem and we believe that, eventually, 
an analytical solution in the limit $N\to\infty$ is possible 
along the same lines followed here.

\appendix

\section{Appendix: Replica calculation for the MG}

Our goal is to compute and characterize the minimum of 
$H_\eta=\ovl{\avg{A}^2}-\eta N G$, with $G$ given by Eq. \req{G},
in $\DN=\{\vp_{i},~i\in\N\}$. 
Considering $H_\eta$ as an Hamiltonian of a statistical 
mechanic's system, this can
be done analyzing the zero temperature limit. First we build the partition
function
\be
Z(\beta)=\Tr_{\pi} e^{-\beta H_\eta\{\pi\}},
\ee
where $\beta$ is the inverse temperature and $\Tr_\pi$ stands for an integral
on $\DN$ (we call simply $\pi$ an element of $\DN$). 
The quantity of interest is then 
\be
\min_{\pi\in\DN} H_\eta\{\pi\}=-\lim_{\beta\to\infty}\beta^{-1}\ln Z(\beta).
\label{limbeta}
\ee
This in principle depends on the specific realization $a_{s,i}^\mu$ of rules
chosen by agents. In practice however, to leading order in $N$, all
realizations of $a_{s,i}^\mu$ yield the same limit, which then coincides with
the average of $\min_{\pi\in\DN} H_\eta\{\pi\}$ over $a_{s,i}^\mu$. 
The average of 
$\ln Z$ over the $a$'s, which we denote by $\avg{\ldots}_a$, 
is reduced to that of moments of $Z$ using the replica trick\cite{MPV}:
\be
\Avg{\ln Z}_a=\lim_{n\to 0}\frac{1}{n}\ln\Avg{Z^n}_a
\label{replicatrick}
\ee
With integer $n$ the calculation of $\Avg{Z^n}_a$ amounts to study $n$ {\em
replicas} of the the same system with the same realization of $a_{s,i}^\mu$.
To do this we introduce a set of dynamical variables $\pi_a\equiv
\{\pi_{s,i,a}\}$ for each replica, which are labeled by the additional 
index $a=1,\ldots,n$. Each replica has its corresponding Hamiltonian,
which we write as $H_\eta^a\{\pi_a\}=\overline{A_a^2}-\eta N G_{a,a}$ 
where $A^\mu_a=\sum_{i\in\N}\vp_{i,a}\cdot\va_i$
and $NG_{a,a}=\sum_i |\pi_{i,a}|^2$ (the reason for this notation shall become
clear later). The set of all dynamical variables
for all replicas is the direct product ${\DN}^n$ of $n$ phase spaces $\DN$.
In order to compute 
the limit $n\to 0$ in Eq. \req{replicatrick} one appeals to analytic
continuation of $\Avg{Z^n}_a$ for real $n$. We give here the
details of the calculation in our specific case. 
More details on the nature of
the method can be found in ref. \cite{MPV}. 
We write
\bea
\Avg{Z^n}&=&\Tr_\pi
\prod_{a=1}^n\prod_{\mu\in\P}
\Avg{e^{-\beta\varrho^\mu[(A^\mu_a)^2-\eta NG_{a,a}]}}_{a}
\nonumber\\
&=&\Tr_\pi
\prod_{\mu\in\P}\prod_{a=1}^n
E_{z_a^\mu}
%\int_{-\infty}^\infty \frac{dz}{\sqrt{2\pi}}
\Avg{e^{-i\sqrt{2\beta\varrho^\mu} z_{a}^{\mu} A^\mu_a}}_{a}
e^{\beta\eta NG_{a,a}}
\label{eq1}
\eea
where $E_z[\ldots]$ stands for the expectation over the Gaussian variable (unit
variance an zero mean) $z$ and we have introduced one such variable $z_a^\mu$
for each $a$ and $\mu$, using the identity $E_z[e^{-ixz}]=e^{-x^2/2}$. In
addition we used the shorthand $\Tr_\pi$ for the integral over ${\DN}^n$.
The average over $a_{s,i}^\mu$ now factorizes
\[
\prod_{a=1}^n\Avg{e^{-i\sqrt{2\beta\varrho^\mu} z_{a}^{\mu} A^\mu_a}}_{a}=
\prod_{i\in\N}\prod_{s=1}^S\Avg{e^{-i\sqrt{2\beta\varrho^\mu} (\sum_a
z_{a}^{\mu}\pi_{s,i}^a) a_{s,i}^\mu}}_a=
\]
and we can explicitly compute it using the distribution \req{probasim}. 
This gives
\[
=\prod_{i\in\N}\prod_{s=1}^S\cos\left[\sqrt{2\beta\varrho^\mu}\sum_{a=1}^n
z_{a}^{\mu}\pi_{s,i}^a\right]\simeq
\prod_{i\in\N}
\exp\left[-\beta\varrho^\mu\sum_{a,b=1}^n
z_{a}^{\mu}z_{b}^{\mu}\sum_{i\in\N}\vp_{i}^a\cdot\vp_{i}^b\right].
\] 
In the last passage we used the relation $\cos x\simeq e^{-x^2/2}$ which is
correct to order $x^2$ in a power expansion. This is justified 
as long as $\varrho^\mu\to 0$ as $P=\alpha N\to\infty$ for
each $\mu\in\P$. Note that, because of this reason, we would have got
the same result for any generic distribution $P_i(a)$ of $a_{s,i}^\mu$
such that $\avg{a}=0$ and $\avg{a^2}=1$. This allows us to understand
why models with continuum strategies $a_{s,i}^\mu\in\rit$, such as the
one proposed in ref. \cite{cavagna2}, yield the
same results as the one with binary strategies, which we are discussing 
here. Before going back to Eq. \req{eq1}, we introduce the matrices
$\hat G\equiv\{G_{a,b},~a,b=1,\ldots,n\}$ and
$\hat r\equiv\{r_{a,b},~a,b=1,\ldots,n\}$ through the identities
\[
1=\!\int\! dG_{a,b} \delta \!\left( G_{a,b}-\!\frac{1}{N}
\sum_{i\in\N}\vp_{i}^a \cdot\vp_{i}^b\right)\propto
\!\int\! dr_{a,b}dG_{a,b}e^{\frac{\alpha\beta^2 r_{a,b}}{2}
\left( \sum_{i}\vp_{i}^a \cdot\vp_{i}^b-N G_{a,b}\right)}
\]
for all $a\ge b$, where $\delta(x)$ is Dirac's delta function and we used its
integral representation. The only part depending on the $\pi_{s,i}^a$ 
in $\avg{Z^n}$ is 
$e^{\alpha\beta^2 \sum_{a\ge b} r_{a,b}\sum_{i}\vp_{i}^a \cdot\vp_{i}^b /2}$.
This can be factorized in the agent's
index $i$ and so the integral $\Tr_\pi$ on ${\DN}^n$ can be factorized into
$N$ integrals over $\Delta^n$ (=the direct product of the simplexes of
the $n$ replicas of the same agent's mixed strategies). 
With this we can write 
\be
\Avg{Z^n}=\!\int\! dr_{a,b}dG_{a,b} e^{-\beta n N F_\beta(\hat G,\hat r)}
\label{saddle}
\ee
where, specializing to the case $\varrho^\mu=1/P$\footnote{A generic 
distribution $\varrho^\mu$ can also be handled, though with heavier 
notations.}, 
\bea
F_\beta(\hat{G},\hat r)&=&\frac{\alpha}{2n\beta}\ln{\rm det}\left[\hat I +
\frac{2\beta}{\alpha}\hat G\right]+
\frac{\alpha\beta}{2n}\sum_{a,b}r_{a,b}G_{a,b}\nonumber\\
&\,&-\frac{1}{n\beta}
\ln\Tr_{\pi\in\Delta^n}
\exp\left[\frac{\alpha\beta^2}{2}\sum_{a,b}r_{a,b}
\vp^a\vp^b
\right]-\eta\sum_a G_{a,a},
\eea
where $\hat I$ is the identity matrix.
The first term arises from the expectation over $z_a^\mu$. This
factorizes for each $\mu$ and one is left with a Gaussian integral
over $\vec z\in \rit^n$. 
The second and the third terms arise from the integral 
representation of the delta functions\footnote{For simplicity we 
have also done the transformation $r_{a,b}\to r_{a,b}/2$ for $a\ne b$ 
so that $\sum_{a\ge b} \to\sum_{a,b}$.}. 

The key point is that, in the limit $N\to\infty$ the integral over the 
matrices $\hat r$ and $\hat G$ in Eq. \req{saddle} are dominated by 
their saddle point value, i.e. by the values of $r_{a,b}$ and $G_{a,b}$ 
for which $F$ attains its
minimum value\footnote{Note that, by Eq. \req{limbeta}, we shall also be
interested in the limit of $\beta\to\infty$ in the end!}. One should then
study the first order conditions $\partial F/\partial r_{a,b}=0$ and $\partial
F/\partial G_{a,b}=0$ for all $a,b$. Here we focus on the {\em replica 
symmetric} approximation where we assume that the matrices 
for which $F$ attains its extreme have the form 
\be
G_{a,b}=g+(G-g)\delta_{a,b},\qquad\qquad
r_{a,b}=r+(R-r)\delta_{a,b}.
\label{RSapprox}
\ee
This ansatz is correct for $\eta\le 0$ and for $\eta>0$ and $\alpha$ large 
enough\cite{andemar}.
The reason for this is that $H_\eta$ is a non-negative definite quadratic
form in $\DN$. Hence it has a very simple {\em energy landscape},
characterized by a single {\em valley}.
Taking the limit $n\to 0$, Eq. \req{replicatrick} then gives 
\bea
%%%%%\frac{\Avg{\ln Z(\beta)}_a}{N}
F^{(RS)}_\beta(Q,q,R,r)&=&\frac{\alpha g}{\alpha+2\beta(G-g)}+
\frac{\alpha}{2\beta}\ln\left[
1+\frac{2\beta(G-g)}{\alpha}\right]-\eta G
\nonumber\\
&\,&+\frac{\alpha\beta}{2}\left(RG-rg\right)-
\frac{1}{\beta}E_{\vz}\!\left\{
\ln\Tr_{\pi}
\exp\left[-\beta V_{\vz}(\vp)\right]\right\}
\label{F}
\eea
where $\Tr_\pi$ is now the integral over the simplex $\Delta$ of a
single agent's mixed strategies and we defined, for convenience, 
the potential
$V_{\vz}(\vp)=\sqrt{\alpha r}\,\vz\cdot\vp-\frac{\alpha}{2}
\beta(R-r)|\pi|^2$. 
The parameters $g,G,r$ and $R$ are fixed by the first order conditions
$\partial F^{(RS)}_\beta/\partial g=0$, 
$\partial F^{(RS)}_\beta/\partial G=0$, 
$\partial F^{(RS)}_\beta/\partial r=0$
and $\partial F^{(RS)}_\beta/\partial R=0$. 
These equations, finally, have to be studied in
the limit $\beta\to\infty$, where one recovers the minimum of $H_\eta$ by Eq.
\req{limbeta}, i.e.
\[
\lim_{N\to\infty}
\min_{\pi\in\DN} \frac{H_\eta\{\pi\}}{N}=-\lim_{\beta\to\infty}
\frac{1}{\beta N}
\Avg{\ln Z(\beta)}=\left.\lim_{\beta\to\infty}
F^{(RS)}_\beta(Q,q,R,r)\right|_{\rm sp}
\]
where the subscript sp means that we compute the function $F^{(RS)}_\beta$ at
the saddle point values of $Q,q,R$ and $r$.

It is convenient to define the parameters 
\be
\chi=\frac{2\beta(G-g)}{\alpha},~~~~~~~y=\frac{\sqrt{g/\alpha}}
{1+\eta(1+\chi)}
\label{params}
\ee
In the limit $\beta\to\infty$, we first look for solutions
where $g\to G$ and $\chi$, which we call {\em susceptibility}, 
remains finite. This implies that two replicas of the same system 
converge in the long run to the same stationary state.
Using the saddle point equations, and $g=G$, we can rewrite 
\be
V_{\vec z}({\vec \pi})=
\frac{2y\,{\vec z}\cdot{\vec \pi}
+\pi^2}{1+\chi},~~~~~~\beta\to\infty
\ee
The last term in Eq. \req{F} is dominated by the mixed 
strategy $\vp^*(\vz)$ which is the solution of
\be
\vp^*(\vz)={\rm arg}\min_{\pi\in\Delta} V_{\vz}(\vp).
\ee
We find that $G=g=E_{\vz}[\vp^*(\vz)]$, which is then 
a function of $y$ only $G\equiv G(y)$. Upon defining 
$\zeta(y)=E_{\vz}[\vz\cdot\vp^*(\vz)]$, we find
\[
\chi(y)=-\frac{\zeta(y)}{\sqrt{\alpha G(y)}+\zeta(y)}
\]
The second of Eqs. \req{params} becomes an equation for 
$y$ as a function of $\alpha$ which has two implicit solutions.
These can be expressed as explicit solutions for $\alpha$ as
a function of $y$ and $\eta$:
\be
\alpha=\frac{1}{G}\left[\frac{G-y\zeta\pm\sqrt{(G+y\zeta)^2-4\eta y \zeta G}}
{2(1-\eta)y}\right]^2
\label{branch}
\ee

\subsection{$\eta\le 0$}

The solutions of Eq. \req{branch}, for $\eta<0$ describe the two 
branches $\alpha<\alpha_c$ and $\alpha>\alpha_c$. In particular 
for $\eta\to 0^-$ these solutions become
\be
\alpha y^2=G(y),~~~~~~~\alpha G(y)=\zeta^2(y).
\label{eqsy}
\ee

Let us discuss first the case $\eta\to 0^-$: 
The {\em free energy} per agent is 
\be
\lim_{N\to\infty}\frac{H}{N}=-\lim_{N\to\infty}
\frac{\Avg{\ln Z(\beta)}_a}{N}=\frac{G}{(1+\chi)^2}
\label{freeenergy}
\ee

These equations are transcendental and we could not find an 
explicit solution for generic $S$. Nevertheless, they represent 
a great simplification with respect to the original problem. 
The main technical difficulty lies in the evaluation of the 
functions $G(y)=E_{\vz}[|\pi^*(\vz)|^2]$ and $\zeta(y)=
E_{\vz}[\vz\cdot\vp^*(\vz)]$, which can be
computed numerically to any desired accuracy $\forall S$. 

The first of Eqs. \req{eqsy} gives the $\alpha>\alpha_c$ phase. 
This solution has $\chi>0$ finite and $H>0$ non-zero. As $\alpha$ 
decreases $\chi$ increases and it diverges as $|\alpha-\alpha_c|^{-1}$
when $\alpha\to\alpha_c^+$. In this limit Eq. \req{freeenergy} implies
that $H\sim |\alpha-\alpha_c|^2$ vanishes. The critical point 
$\alpha_c=\alpha(y_c)$ is obtained imposing $\chi=\infty$, which 
gives $G(y_c)=-y_c\zeta(y_c)$. By the numerical evaluation of
the functions $G(y)$ and $\zeta(y)$, we find
\be
\alpha_c(S)\cong\alpha_c(2)+\frac{S-2}{2}
\ee
to a high degree of accuracy. It might be that this equation
is exact but we could not prove it. An interesting relation for 
$\alpha_c(S)$ can be derived by algebraic considerations: 
Note that for each $\pi_{s,i}>0$ the equation
\be
\frac{\partial H}{\partial \pi_{s,i}}=2\sum_{j,s'}\ovl{a_{s,i}a_{s',j}}
\pi_{s',j}=0 
\label{dHdpi}
\ee
must hold. This is a set of linear equations in the variables 
$\pi_{s,i}>0$. The $NS\times NS$ matrix $\ovl{a_{s,i}a_{s',j}}$
is built with $P$ dimensional vectors $a^\mu_{s,i}$ and therefore
has at most rank $P$. In other words there are only $P$ independent 
equations \req{dHdpi}. In addition there are $N$ normalization
conditions on $\pi_{s,i}$.
The system becomes dynamically degenerate when the number of free 
variables $\pi_{s,i}$ becomes bigger than the number $P+N$ of
independent equations and, exactly at $\alpha_c$ the two are equal. 
Dividing this condition by $N$ gives the desired equation
\be
\sum_{s=1}^S E_{\vz}\{\theta[\pi^*_s(\vz)]\}=
%SE_{\vz}\{\theta[\pi^*_1(\vz)]\}=
\alpha_c(S)+1.
\ee
The left hand side is the average number of strategies used 
by agents (called $n$ in section \ref{agentview}). 
Note that this equation implies that $\alpha_c(S)$
cannot grow faster than linear in $S$. Also $\alpha_c(S)\propto S/2$
imply that agents use on average $1/2$ of their strategies at
$\alpha_c$.

The second of Eqs. \req{eqsy} gives the $\alpha<\alpha_c$ phase.
Note indeed that with this choice $\chi\simeq -1/\eta
\to\infty$ and $H\sim \eta^2\to 0$ as $\eta\to 0^-$. 
At odds with the solution for $\alpha>\alpha_c$,
this equation only arises if $\eta< 0$ and in the limit
$\eta\to 0^-$. 
With $\eta=0$ the saddle point 
equations have only a solution with $G>g$ in the limit $\beta\to
\infty$. This is because for $\alpha<\alpha_c$ the set ${\cal M}$
where $H=0$ is not a single point, but rather a connected set.
The replica method with $\eta=0$ takes an average
on all the set ${\cal M}$ and so it gives results which are not
representative of a particular system\footnote{Note indeed that $g$ 
has the interpretation of the overlap between two replicas of the
same system, so that $g<G$ means that the two replicas are not
identical.}. In order to select a single point in ${\cal M}$ one
may consider the limit $\eta\to 0^-$. Since the term $-\eta\,N\,G$ in 
$H_\eta$ breaks the degeneracy of equilibria for $\eta=0$, 
the limit $\eta\to 0^-$ selects the equilibrium which 
is closest to the random initial condition $\pi_{s,i}(0)=1/S$ for
all $i\in \N$ and $s=1,\ldots,S$. This describes the stationary state 
of a system of agents with no prior beliefs ($U_{s,i}(t=0)=0$, 
$\forall s,i$). 

In both phases, once the saddle point equations are solved,
one can derive the full statistical characterization of the 
system. For instance the fraction of agents playing a 
strategies in a neighborhood $d\vp$ of $\vp$ is given by
$p(\vp)d\vp=E_{\vz}[\delta(\vp^*(\vz)-\vp)]d\vp$.

\subsection{$\eta>0$}

Let us for simplicity consider the simpler case $S=2$. 
The solution with $G=g$ exists for $\alpha\ge 1/\pi$.
For $\alpha>[\pi(1-\eta)^2]^{-1}$ this solution has $G=g<1$, which
means that agents do not all play pure strategies. When 
$\alpha\to [\pi(1-\eta)^2]^{-1}$, $G\to 1$ and the solution becomes
independent of $\eta$. In other words, the solution merges with
the solution for $\eta=1$. In its turn this solution breaks down,
with $\chi\to\infty$ and $H_1/N=\sigma^2/N\to 0$ when $\alpha\to 1/\pi$.
Below this point, only solutions with $G<g$ and $H_1/N=0$ exist. This 
behavior is well documented in figure \ref{figsigvseta}. However, for
$\eta>0$, one needs to go beyond the 
simple approximation for $G_{a,b}$ and $r_{a,b}$ in Eq. \req{RSapprox}.
Therefore we shall refrain from a more detailed discussion and rather
refer the interested reader to a forthcoming publication \cite{andemar}.

\end{document}